\documentclass[iop]{emulateapj-rtx4}

\usepackage{longtable}
\usepackage{natbib}

\newcommand{\hip}{\textit{Hipparcos}\ }
\newcommand{\msun}{M_{\odot}}
\newcommand{\rsun}{R_{\odot}}
\newcommand{\mjup}{M_{JUP}}

\newcommand{\mps}{{\rm m\ s}^{-1}}
\newcommand{\kmps}{{\rm km\ s}^{-1}}
\newcommand{\msini}{M_p \sin i}
\newcommand{\vsini}{v \sin i}
\newcommand{\feh}{\rm [Fe/H]\ }
\newcommand{\logg}{\log \ g}
\newcommand{\iodine}{\rm I_2}

\shorttitle{Five New Exoplanets}
\shortauthors{Harakawa et al.}

\begin{document}


\title{Five New Exoplanets Orbiting Three Metal-Rich, Massive Stars: Two-Planet Systems Including Long-Period Planets, and an Eccentric Planet}


\author{Hiroki Harakawa\altaffilmark{1}, Bun'ei Sato\altaffilmark{2}, 
Masashi Omiya\altaffilmark{2}, Debra A. Fischer\altaffilmark{3}, 
Yasunori Hori\altaffilmark{4}, Shigeru Ida\altaffilmark{2},
Eiji Kambe\altaffilmark{5}, Michitoshi Yoshida\altaffilmark{6},
Hideyuki Izumiura\altaffilmark{5}, Hisashi Koyano\altaffilmark{5},
Shogo Nagayama\altaffilmark{1}, Yasuhiro Shimizu\altaffilmark{5},
Norio Okada\altaffilmark{3}, Kiichi Okita\altaffilmark{5},
Akihiro Sakamoto\altaffilmark{5},
and Tomoyasu Yamamuro\altaffilmark{7}
}
%\affil{}
\email{h.harakawa@nao.ac.jp}

%\and

%\author{}


\altaffiltext{1}{National Astronomical Observatory of Japan, 
2-21-1 Osawa, Mitaka, Tokyo 181-8588, Japan}
\altaffiltext{2}{Department of Earth and Planetary Sciences, 
Tokyo Institute of Technology, 2-12-1 Ookayama, Meguro-ku, Tokyo 152-8551, Japan}
\altaffiltext{3}{Department and Astronomy, Yale University, New Haven, CT, 06520, USA}
\altaffiltext{4}{UC Santa Cruz, 1156 High Street, Santa Cruz, CA 95064, USA}
\altaffiltext{5}{Okayama Astrophysical Observatory, 
National Astronomical Observatory of Japan, 
3037-5 Honjo, Kamogata, Asakuchi, Okayama 719-0232, Japan}
\altaffiltext{6}{Hiroshima Astrophysical Science Center, Hiroshima University,
1-3-1 Kagamiyama, Higashi-Hiroshima, Hiroshima 739-8526, Japan}
\altaffiltext{7}{OptCraft, 3-6-18 Higashi-Hashimoto, Midori-ku, Sagamihara 252-0144, Japan}


\begin{abstract}
We report detections of new exoplanets from a radial velocity (RV) survey of metal-rich FGK stars by using three telescopes.
By optimizing our RV analysis method to long time-baseline observations, we have succeeded in detecting five new Jovian-planets around three metal-rich stars HD 1605, HD 1666, and HD 67087 with the masses of $1.3\msun$, $1.5\msun$, and $1.4\msun$, respectively.
A K1 subgiant star HD 1605 hosts two planetary companions with the minimum masses of $\msini = 0.96\mjup$ and $3.5\mjup$ in circular orbits with the planets' periods $P = 577.9$ days and $2111$ days, respectively.
HD 1605 shows a significant linear trend in RVs.
Such a system consisting of Jovian planets in circular orbits has rarely been found and thus HD 1605 should be an important example of a multi-planetary system that is likely unperturbed by planet-planet interactions.
HD 1666 is a F7 main sequence star which hosts an eccentric and massive planet of $\msini = 6.4\mjup$ in the orbit with $a_{\rm p} = 0.94$ AU and an eccentricity $e=0.63$.
Such an eccentric and massive planet can be explained as a result of planet-planet interactions among Jovian planets.
While we have found the large residuals of $\mathrm{rms} = 35.6\ \mps$, the periodogram analysis does not support any additional periodicities.
Finally, HD 67087 hosts two planets of $\msini = 3.1\mjup$ and $4.9\mjup$ in orbits with $P=352.2$ days and $2374$ days, and $e=0.17$ and $0.76$, respectively.
Although the current RVs do not lead to accurate determinations of its orbit and mass, HD 67087 c can be one of the most eccentric planets ever discovered in multiple systems.
\end{abstract}


\keywords{planetary systems --- stars: individual(HD 1605, HD 1666, HD 67087) --- techniques: radial velocities --- techniques: spectroscopic}



\section{Introduction}
Since 1995, precise radial-velocity (RV) searches for exoplanets have unveiled hundreds of planets around solar-type stars.
The improvements of spectrographs and RV analysis techniques have enabled us to detect low-mass planets such as Earth-like planets.
The current detection limit on planetary masses for RV searches is reaching the level of several Earth-masses within 0.1AU around solar-type stars 
\citep{2010Sci...330..653H,2011arXiv1109.2497M}.
To date, masses of known exoplanets range from about one Earth-mass to tens of Jupiter-masses.
Thanks to increasing the number of planet samples, statistical studies become available.
In particular, Jovian-planet samples from 20-year-long surveys have been contributing to reveal their statistical properties over a wide range of orbital distance.
The pictures of the detected exoplanets have been contributing to constrain planet formation theories taking account of evolutions of protoplanetary disks and planet's orbits such as planetary migration and planet-planet scattering
\citep[e.g. population synthesis: ][]{2004ApJ...604..388I,2013ApJ...775...42I,
2009A&A...501.1139M}, 
however, statistical studies for the planet distribution
\citep[e.g.][]{1999ApJ...526..890C,2008PASP..120..531C,
2010ApJ...709..396B,2012A&A...543A..45M,2013A&A...560A..51A} 
are still open to debate, particularly on the correlation with metallicities of their host stars.
Recently, near-infrared observations of disk fractions in young clusters suggested short lifetimes of protoplanetary disks around low-metallicity stars \citep{2010ApJ...723L.113Y}. 
Theoretical studies on disk evolution under X-ray photoevaporation supported that disk lifetimes may be positively correlated with the stellar metallicity 
\citep[e.g.][]{2010MNRAS.402.2735E}.
The disk metallicity also determines the total amount of building materials for planets.
These imply that a long-lived disk in a metal-rich environment may enhance the occurrence rate of Jovian planets such as hot-Jupiters and Jupiter-analogs.
While previous studies of the planet-metallicity correlation
\citep[e.g.][]{1997MNRAS.285..403G,2003A&A...398..363S,2005ApJ...622.1102F,2010PASP..122..905J}
have mainly focused on the occurrence rate of Jovian planets integrated over planetary mass and orbital period, it is also important to reveal dependence of the mass-period distribution of planets on stellar metallicity for an in-depth understanding of orbital migration and formation of giant planets in various environments.

Since 2009, we have conducted RV surveys for metal-rich solar-type stars mainly using the 188 cm telescope at Okayama Astrophysical Observatory (OAO) and the Subaru telescope.
This planet-search program originates from the N2K project, which was an international collaboration between American, Japanese and Chilean astronomers aiming at finding hot-Jupiters, especially transiting ones, around metal-rich FGK stars using large aperture telescope such as the Keck, Subaru, and Magellan
\citep[e.g.][]{2005ApJ...620..481F,2005ApJ...633..465S,2008AJ....136.1901L,2012ApJ...744....4G}.
As for the Subaru telescope part, in addition to the detections of transiting HJs, continuous observations revealed Jovian planets from the including those in years-long orbits
\citep{2007ApJ...669.1336F,2009PASP..121..613P,2012ApJ...744....4G}.
The metal-rich nature of the N2K targets totally fits our purpose which aims at unveiling orbital distribution of giant planets around metal-rich stars over a wide range of orbital periods.
In order to achieve this objective, long time baseline RV observations are of particular importance in revealing a population of planets at distant orbits where Jupiter-analogs are thought to have formed via a core-accretion process.
Continuous RV observations also lead to secure detections of multiple systems and eccentric planets 
\citep[e.g.][]{2009ApJ...693.1084W, 2014ApJ...783..103W} 
which will provide useful information about orbital evolutions such as an orbital migration which occurs due to an interaction between the disk-gas and planetesimals, and gravitational interactions among planetary/stellar companions
\citep[e.g.][]{2008ApJ...678..498N,2011ApJ...742...72N}.

In this paper, we report the detections of five new planets around three metal-rich stars HD 1605, HD 1666, and HD 67087 from observations at telescopes of OAO, Subaru, and Keck.
In Section \ref{sec_obs}, we present stellar characteristics and orbital solutions for the three stars.
We also briefly describe the method for RV measurements including our improvements in RV stability and precision for the Subaru telescope.
In Section \ref{sec_result}, we show our results of planet detections.
We discuss and summarize the detections from a view point of planet formation and evolution in Section \ref{sec_disc}.

\section{Observations}
\label{sec_obs}
\subsection{Targets}
The stellar targets of our long-term planet-search program at OAO and the Subaru telescope are based on those of the N2K project.
More than 14,000 main-sequence and subgiant stars were selected from the \hip catalog 
\citep{1997ESASP1200.....P} 
as potential targets of the N2K project, and their properties had been investigated such as $B-V$ color magnitudes, $JHK$-band photometry from the 2MASS catalogue, stellar luminosity, parallaxes, proper motions, photometric variations, and the existence of stellar companions.
All the targets had been selected with following criteria: (1) $0.4 < B-V < 1.2$, (2) $V < 10.5$ and (3) stars closer than 110 pc.
The approximate 2,000 stars had been selected from the potential targets as the original N2K targets by setting metallicity criterion of $\feh > +0.1$ estimated from photometry.

As the component of the N2K project, the Subaru telescope had observed a total of 635 stars out of the original N2K targets at least once as of 2009, and we adopted the 635 stars as our primary targets for the subsequent observations.
Among them, we have intensively observed 50 bright stars at OAO in order to confirm planet candidates. Also, we have observed the 635 primary targets evenly at the Subaru telescope in order to mainly extend the time baseline of our samples for statistical studies.

\subsection{OAO/HIDES Observations}
For observations at OAO, we use the 188 cm telescope and the HIgh Dispersion Echelle Spectrograph
\citep[HIDES; ][]{1999oaaf.conf...77I} 
installed at its coud\'{e} focus.
Until December 2010, we used the conventional slit aperture of HIDES (HIDES-S).
The slit width was set to ${\rm 250 \mu m}(0''.95)$, giving a spectral resolution $R = 50000$ by 4 pixel sampling. A mosaic of three CCDs is capable of a simultaneous wavelength coverage from 3700 to 7500 \AA.
For precise RV measurements, we used an iodine absorption cell 
\citep[$\iodine$ cell; ][]{2002PASJ...54..865K}.
For $V\sim8$ stars with an exposure time of 30 minutes, a typical signal-to-noise ratio ${\rm SNR} > 130\ {\rm pix}^{-1}$ at $5500$ \AA,with which we can
achieve the RV precision of $7\ \mps$.
After January 2011, we have been using the HIDES fiber-feeding system (HIDES-F) that has been newly developed for attaining better throughput \citep{2013PASJ...65...15K}.
The HIDES-F uses an image-slicer as an entrance aperture of the spectrograph and its spectral resolution is fixed to $R=47,000$.
We can typically obtain spectra with two times better efficiency (${\rm SNR} > 180\ {\rm pix}^{-1}$ at 5500 \AA\ for $V\sim8$ stars with a 30 min-exposure) compared to the previous slit-observations, with which the RV precision can reach down to $3\ \mps$.
We have been observing several RV standard stars using both a slit and image-slicer, and monitoring these RV variations.
As a Doppler-shift standard, we applied a newly obtained template-spectrum of HDS whose details are described in section \ref{sec:obs_sub}.
The calculated RVs did not show significant systematics between the slit-data and image-slicer-data within the typical rms (root-mean-square) of $3\ \mps$.
Thus, we are able to consider that there is no significant RV systematics between the two settings of HIDES.

Since the N2K project supposed planet searches by 10 m-class telescopes, the apparent magnitude of the targets tends to be fainter than 9 in $V$-magnitude.
Stars with magnitudes of $V > 9.5$ are too faint for precise RV measurements by 2 m-class telescopes.
Thus, we have selected 50 brightest stars that have magnitudes of $7.5<V<9.5$ from our primary targets.
Almost all of the 50 targets show significant RV variations of $\sigma_{\rm RV}>20\ \mps$, but their periodicities have not been confirmed yet.
Such RV variations are comparable to a RV amplitudes caused by a giant planet orbiting around a solar-mass star with a semi-major axis of 1 AU.
We obtained stellar parameters of the 50 targets from the Spectroscopy Made Easy 
\citep[SME; ][]{1996A&AS..118..595V,2005ApJS..159..141V} analysis 
and the \hip catalogue.

\subsection{Subaru/HDS Observations}
\label{sec:obs_sub}
From 2004 to 2010, the original N2K survey collected several RV data for each of our primary targets.
Since 2011, we have conducted follow-up observations for a part of them using the Subaru telescope.
While observations at OAO mainly focused on planet-host candidates which showed significant RV variations.
In order to obtain reliable statistical results, we need to gather RV data without selection biases.
Thus, we collected RV data uniformly from the primary targets as well as planet-host candidates, using the Subaru telescope.

We utilize the High Dispersion Spectrograph 
\citep[HDS;][]{2002PASJ...54..855N} 
on the 8.2-m Subaru telescope.
We mainly adopted the set-up of StdI2b, which simultaneously covers a wavelength region from 3500 to 6100\AA\ by a mosaic of two CCDs.
The slit width was set to $0''.8$ for the first 2 year-observations (2004--2005) and $0''.6$ for the later observations until 2011, giving a spectral resolution of $R=45000$ and 60000, respectively.
In 2012, we used a newly-developed image-slicer (IS\#1) with the resolution of $R=110000$ in order to improve the throughput of each exposure
\citep[see][for details of the image-slicer]{2012PASJ...64...77T}.
Our set-up allows us to obtain ${\rm SNR} \sim 150\ {\rm pix}^{-1}$ at 5500 \AA\ with 30--300 sec exposure times for our typical targets with $V \sim 8.5$.
We used an $\iodine$ cell to provide a fiducial wavelength reference for precise RV measurements 
\citep{2002PASJ...54..865K,2002PASJ...54..873S}.
Although the $\iodine$ cell was originally installed just behind the entrance slit of the spectrograph, it was moved in front of it in 2006.

We basically used an analysis code described in 
\cite{2002PASJ...54..873S}
for precise RV measurements which generates the stellar template from star+$\iodine$ spectra.
Stellar templates obtained in the early phase of the N2K survey sometimes cause poor RV accuracy for years-long time-baseline observations with HDS.
We also conducted additional observations of the N2K targets in order to refine stellar templates.
The observational setting required for the stellar templates is almost the same as that for RV measurements, except for using extremely high resolution of $R=160000$ (the slit width of $0''.2$ without an $\iodine$ cell).
The point-spread-functions (i.e. instrumental-profiles) are estimated from an $\iodine$-superposed flat spectrum that was observed at the nearest time before/after the exposure.
Using the newly-obtained stellar templates, we can achieve the rms of RVs lower than $3\ \mps$ with about seven-year-long stability.
We optimized our analysis pipeline for this refined template, and applied it to RV measurements for HIDES data as a common reference of Doppler-shifts.

In order to check the long-term RV stability using the whole data from HIDES and HDS, we have been observing several RV standard stars and known planet hosts such as HD 38801 
\citep{2010ApJ...715..550H}
using the two spectrographs.
According to our analysis for "slit-IS" combined set of data from HDS, it has been shown that there is no significant systematic between slit data and image-slicer data.
We have also confirmed that the I2-cell position against the entrance of HDS does not influence RV systematics significantly.
After these checks, we have performed the RV measurements for all the HIDES and HDS data and confirmed that the rms of RVs for the standard stars have been achieved 4--5 $\mps$.
Thus, we can consider that no significant RV-offset between the two sites was detected with a mixture of slit-data and image-slicer-data.

\subsection{Keck/HIRES Observations}
Between 2006 January and 2010 September, we also used the Keck telescope with HIRES spectrograph 
\citep{1994SPIE.2198..362V} 
to follow up some promising candidates of planet-host stars.
The Keck observations were obtained using an $\iodine$ cell and typically with a slit width of 0''.86 corresponding to a spectral resolution of $R=65000$.
The exposure time is optimized to achieve the typical signal to noise ratio of 150--240 per pixel.
The mean RV precision of $1.5 \mps$ was yielded based on the RV analysis method proposed by 
\cite{1992PASP..104..270M}, and \cite{1996PASP..108..500B}.
Since the RVs of HIRES were calculated individually from those of HIDES and HDS, the RV offsets between the two datasets are fitted as free parameters.

\section{Stellar Characteristics and Orbital Solutions}
\label{sec_result}
\subsection{HD 1605}
\label{subsec_1605}
HD 1605 (HIP1640) is a K1 subgiant star with $V=7.52$ and a color index of $B-V=0.96$.
The \hip parallax of 11.82 mas corresponds to a distance of 84.6 pc, i.e., the absolute V magnitude of $M_V=2.88$.
The SME analysis yields the stellar parameters of $T_{\rm eff}=4757\pm50$ K, $\vsini = 0.54\pm0.5\ \kmps$ [Fe/H]$=+0.21\pm0.05$, and $\logg = 3.40\pm0.08$.
Adopting the determined $T_{\rm eff}$, $B-V$, $M_V$, and bolometric collection of $B.C.=-0.432$ \citep{1996ApJ...469..355F}, we have estimated the bolometric luminosity to be $L=6.6\pm0.76L_{\odot}$, and stellar radius of $R_{*}=3.8\pm0.4 \rsun$ based on the Stefan-Boltzmann law.
We estimated the stellar mass from the best-fit interpolation of stellar evolution tracks calculated by 
\cite{2002A&A...391..195G}
to be $M_{*}=1.31\pm0.11M_{\odot}$ with its age of $4.59\pm1.37$ Gyr.
In fact, [Fe/H]$>+0.2$ is out of range of available data of stellar evolution models by 
\cite{2002A&A...391..195G}.
Thus, we have extrapolated the existing tracks for models with a higher stellar metallicity
\citep[see][for the treatment of our extrapolation method]{2012PASJ...64...34O}.
All the stellar parameters are listed in Table \ref{tbl_str}.

We made the first observations for HD 1605 at the Subaru telescope and detected large RV variations. 
After that, the star has been monitored using HDS, HIDES and HIRES for about 9 years (from December 2004 to August 2013).
We excluded data with low SNR under bad weather conditions, and then obtained 75 RV data, 14 and 61 of which were observed by HDS and HIDES, respectively.
Combining 17 RV data from HIRES, we have obtained 92 RV data in total for HD 1605.
The observation dates, RVs, internal errors, and observation sites are listed in table \ref{tbl_obs1605}.

We determined parameter values of the Keplerian orbit by minimizing the $\chi^2$ statistic.
We adopted the AMOEBA algorithm 
\citep{Nelder1965Simplex,1986nras.book.....P}
which is well-suited to the high-dimensional parameter spaces.
The definition of $\chi^2$ is written as
\begin{equation}
\chi^2 = \sum_{i}^{N_{\rm site}} \sum_{j}^{N_i,{\rm obs}} 
\frac{1}{N_{\rm total} - N_{\rm dim}} 
\left(
\frac{v_{ij,{\rm obs}} - v_{ij,{\rm calc}} + \gamma_i}
{\sigma_{ij,{\rm error}}}
\right)^2,
\end{equation}
where $N_{\rm site}$, $N_{i,{\rm obs}}$, $N_{\rm total}$, $N_{\rm dim}$,
$v_{ij,{\rm obs}}$, $\sigma_{ij,{\rm error}}$, $v_{ij,{\rm calc}}$, and $\gamma_i$ 
are the number of observational sites,
the number of observations on the $i$-th site, 
the number of total observations, 
the degree of freedom on the fitting, 
the $j$-th observed RV and observational uncertainty from each site,
the calculated RVs from the model-orbit corresponding to $j$-th observation time on $i$-th site,
and RV offsets on the $i$-th site, respectively.
In fact, the number of observational sites can be written as the number of stellar templates.
A quantity of the RV offset depends on the time when the stellar template was observed.
In this work, we used not more than two stellar templates in total that had been obtained from HDS and HIRES.
Therefore, the appropriate number of site is $N_\textrm{site} = 1$ or $2$.
Each template was applied to RV measurements of the HDS-HIDES combined data set, and HIRES data set, respectively.

\begin{figure}
\epsscale{.65}
\plotone{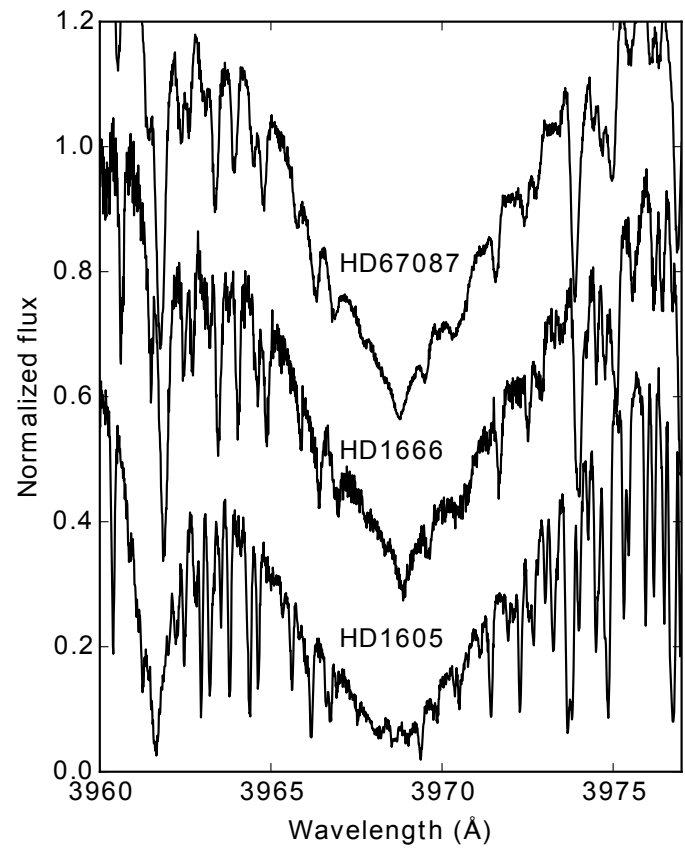}
\caption{Spectra centered on the Ca II H line for the three planet-host stars. All of the stars show no significant emission in the core of lines indicating chromospherically quiet. An appropriate offset is added to each spectrum for clarity.
\label{fig_ca_h}
}
\end{figure}

In order to identify the stellar chromospheric activity, we graphically checked the core emission strength of Ca II HK lines for each star.
As shown in Figure \ref{fig_ca_h}, all of the stars are basically inactive.
We assumed an intrinsic RV noise due to the chromospheric activity of a central star (e.g. starspots and plage) and its surface pulsation, the so-called ``stellar jitter'', to be 6 $\mps$ for ordinary inactive-subgiants \citep{2005PASP..117..657W}, which is added in quadrature 
($\sigma_{\rm error}^2 = \sigma_{\rm obs}^2 + \sigma_{\rm jitter}^2$) to the formal velocity uncertainties for HD 1605 when we fit a Keplerian orbit to the RV data.
At first, we performed the periodogram analysis 
\citep{1976Ap&SS..39..447L,1982ApJ...263..835S} 
and found significant periodicity at $\simeq 2000$ days.
After detrending the 2000 days signal, an another periodic signal appeared at several hundred days period.
Although we considered a two-planet model for the Keplerian fitting, its residuals showed a linear slope over the entire observations.
Therefore, we derived the best-fit solution of a two-planet model with a linear trend.
The rms to the fit is reduced from $10\ \mps$ to $6.4\ \mps$.
The reduced chi-square has been improved from 1.9 (2 planet-model) to 0.94 (2 planet-model + linear trend), indicating that the latter model should be more appropriate within our achievable accuracy.
We calculated the errors in orbital parameters at a $68.27\%$ confidence interval (from bottom $15.87\%$ to top $84.13\%$) of the distributions via the bootstrap Monte-Carlo approach.
The RV residuals to the best-fit Keplerian curve are scrambled and added back to the original measurements, and then, we re-fit and derive new parameters.
We achieved $1\sigma$ confidence levels of each parameter by 10000 times iterations.
We also applied the same technique to two other systems reported in the subsequent sections.

\begin{figure}
\epsscale{.65}
\plotone{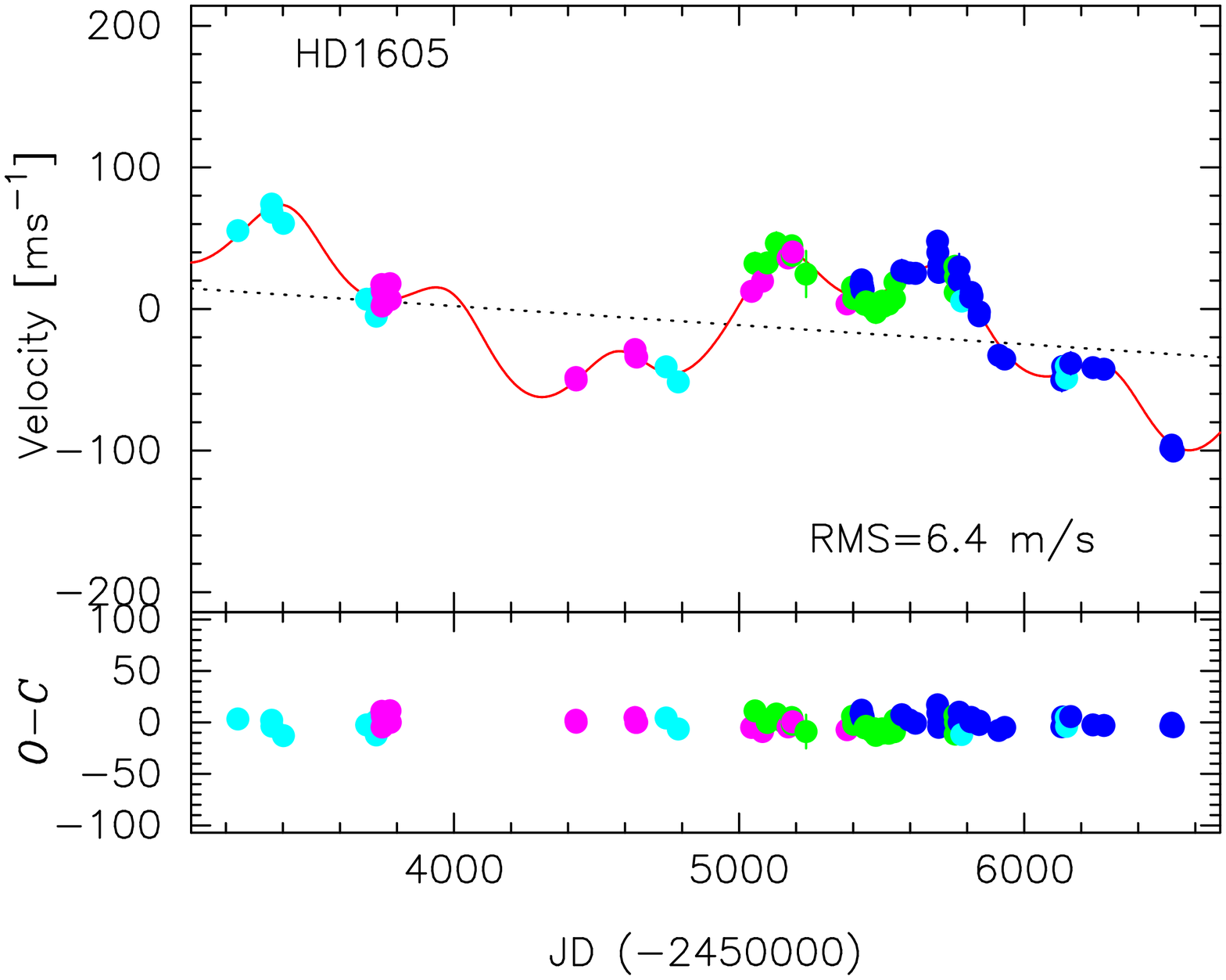}
\plotone{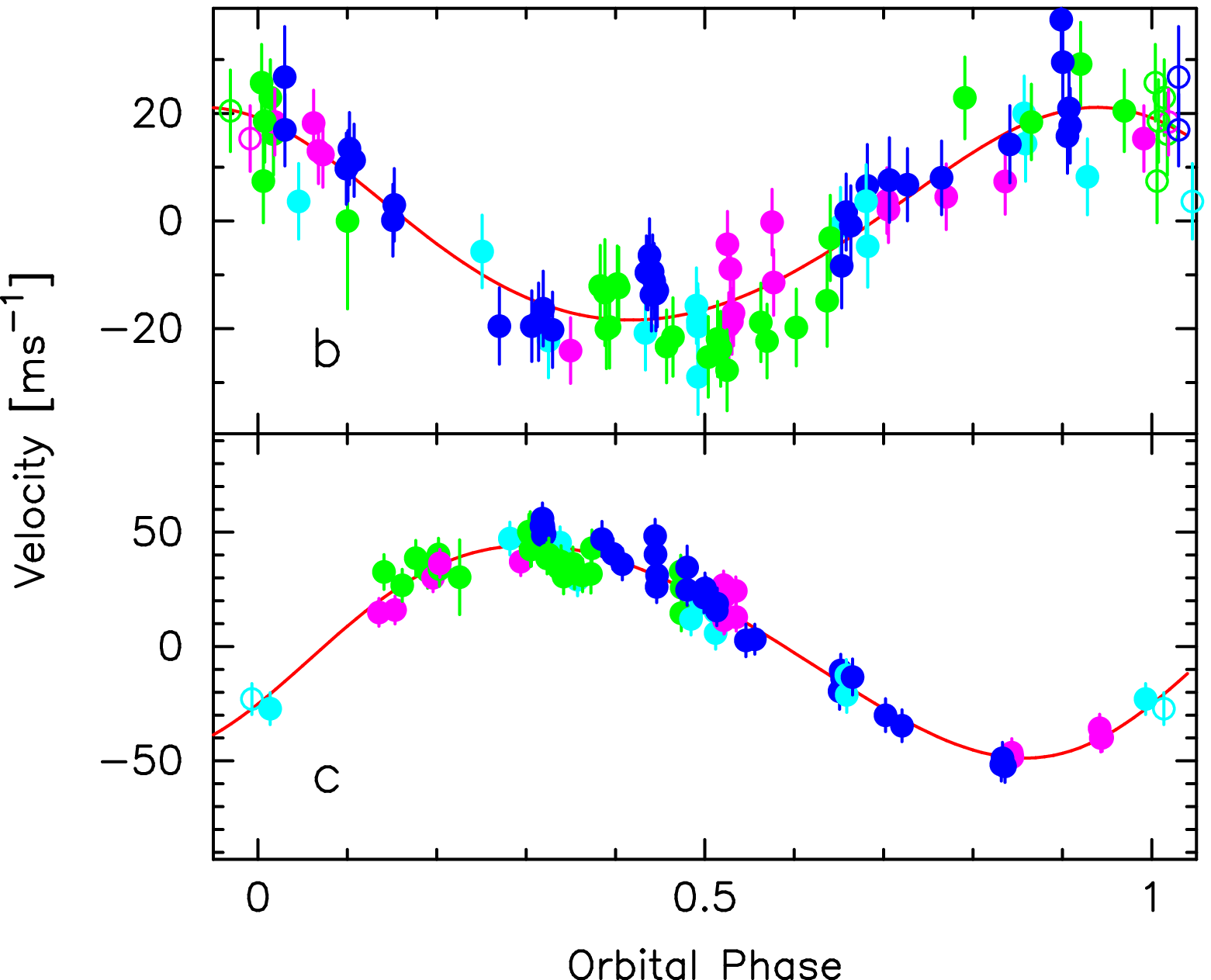}
\caption{
\textit{Top panel:} Observed RV variations and the best-fit orbital solution of HD 1605
with its residuals to the fit.
\textit{Light-blue} points,  
\textit{green} points,
\textit{blue} points,
and \textit{pink} points
represent data from the Subaru/HDS, OAO/HIDES-S, OAO/HIDES-F, and the Keck/HIRES, respectively. 
The best-fit stellar RV motion has been shown as 
\textit{red} solid line.
The error bars are shown as the summations of quadrature of observational errors and
the stellar jitter of $\sigma_{jitter}=6\ \mps$.
The dashed line shows a linear trend that indicates the existence of uncovered 
(sub)stellar or planetary companion(s).
The offsets of $6.7\ \mps$ and $0.1\ \mps$ were introduced to HDS and HIDES data, and HIRES data, respectively.
\textit{Bottom panel:} Phase-folded RV variations for HD 1605.
\label{fig_fit1605}
}
\end{figure}

Figure \ref{fig_fit1605} show the RV variations and the RV curve based on best-fit values.
The best-fit orbital solutions and their errors are listed in table \ref{tbl_prm}.
The orbital solutions of planet ``b'' are $P=577.9^{+5.6}_{-4.9}$ days and the RV semi-amplitude $K_1 = 19.8^{+1.2}_{-0.8}\mps$, corresponding to the minimum planetary mass $\msini = 0.96\mjup$ at the semimajor axis $a_{\rm p} = 1.48$ AU, and those of outer planet ``c'' are $P=2111^{+42}_{-32}$ days and $K_1=46.5^{+1.5}_{-1.2}\mps$, respectively, corresponding to $\msini = 3.48\mjup$ at $a_{\rm p}=3.52$ AU.
The remaining RV trend shows the linear RV variation of $-4.92\ {\mps\ {\rm yr^{-1}}}$.
The orbital period ratio of the planets is not any integer even considering their $1\sigma$ uncertainties.
This suggests that two Jovian planets are not trapped into any mean motion resonance.
Both of the two planets have almost circular orbits ($e<0.1$), and thus we cannot strictly constrain their eccentricities and arguments of periapsis.

\subsection{HD 1666}
HD 1666 (HIP1666) is listed as a F7V star with a visual magnitude of $V=8.17$, and a color index $B-V = 0.53$ in the \hip catalogue.
The \hip parallax of 9.22 mas corresponds to a distance of 108.5 pc, and the absolute visual magnitude is estimated to $M_V=2.76$.
We performed the SME analysis in order to determine the stellar parameters.
Following the method described in 
\cite{2005ApJS..159..141V}, 
we derived an effective temperature $T_{eff}=6317\pm44$K, rotational velocity $\vsini = 5.6\pm0.5 \kmps$, $\logg = 4.06\pm0.06$, and metallicity [Fe/H]$=+0.37\pm0.04$ for the star.
The bolometric luminosity of $L_* = 5.37\pm1.18 L_{\odot}$ was estimated from the $M_V$ and a bolometric correction of $B.C. = -0.016$.
We derived a stellar radius of $R_*=1.93\pm0.49 \rsun$ from the Stefan-Boltzmann law. We performed the isochrone analysis as mentioned in Section \ref{subsec_1605} based on the stellar metallicity, bolometric luminosity and effective temperature.
Using the spectroscopic parameters, we found the mass of $1.50\pm0.07M_{\odot}$ and the age of $1.76\pm0.20$ Gyr for HD 1666.

After the detection of large RV variations for HD 1666 by HDS, we have monitored the star for about 9 years (from December 2004 to August 2013) using the three-spectrographs.
We excluded data with extremely-low SNRs due to bad weather conditions and obtained 99 RV measurements in total, 11, 21, 46, and 21 of which were observed by HDS, HIDES-S, HIDES-F, and HIRES, respectively.
Table \ref{tbl_obs1666} summarizes the observation dates, RVs, internal errors, and observatory sites.

\begin{figure}
\epsscale{.65}
\plotone{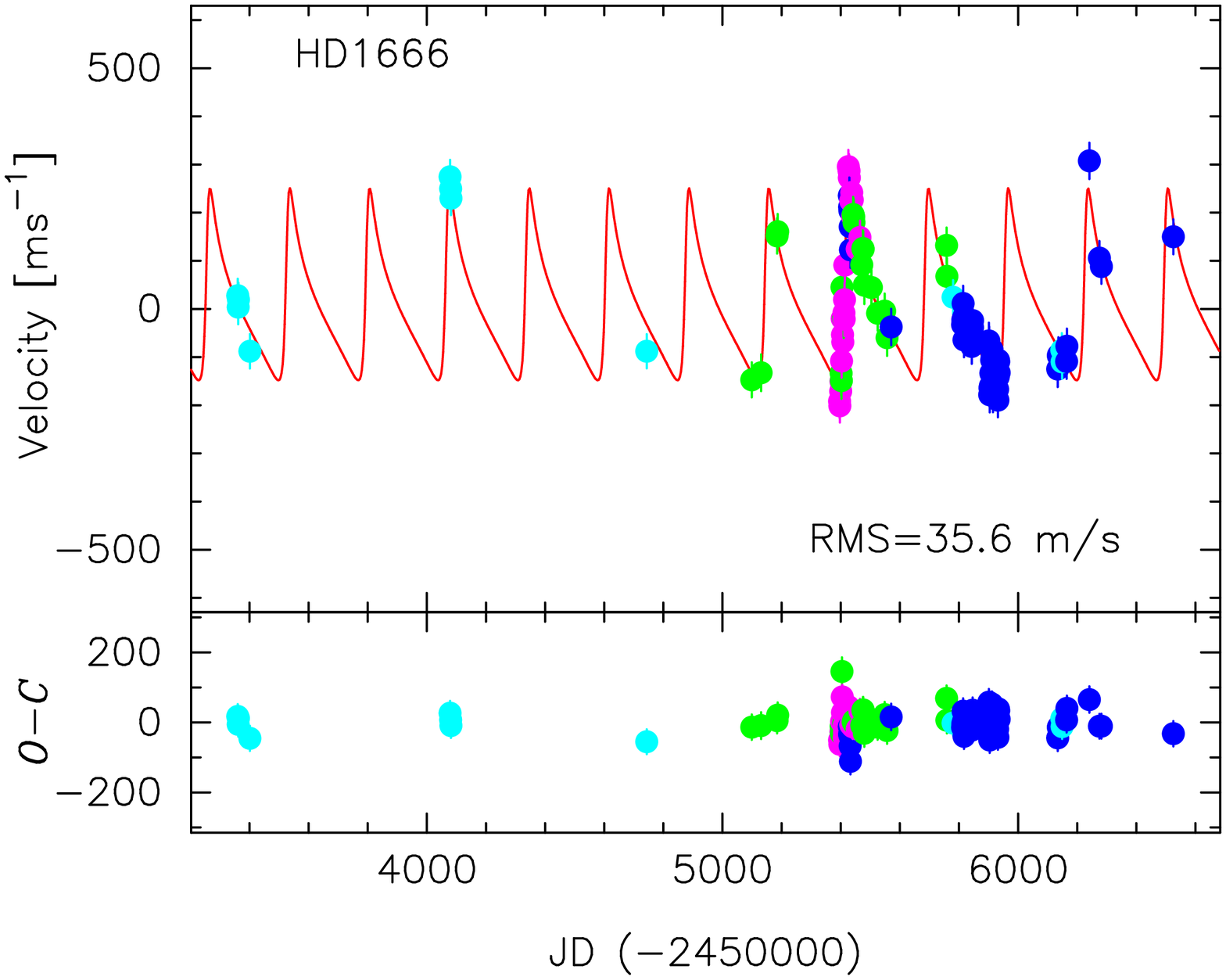}
\plotone{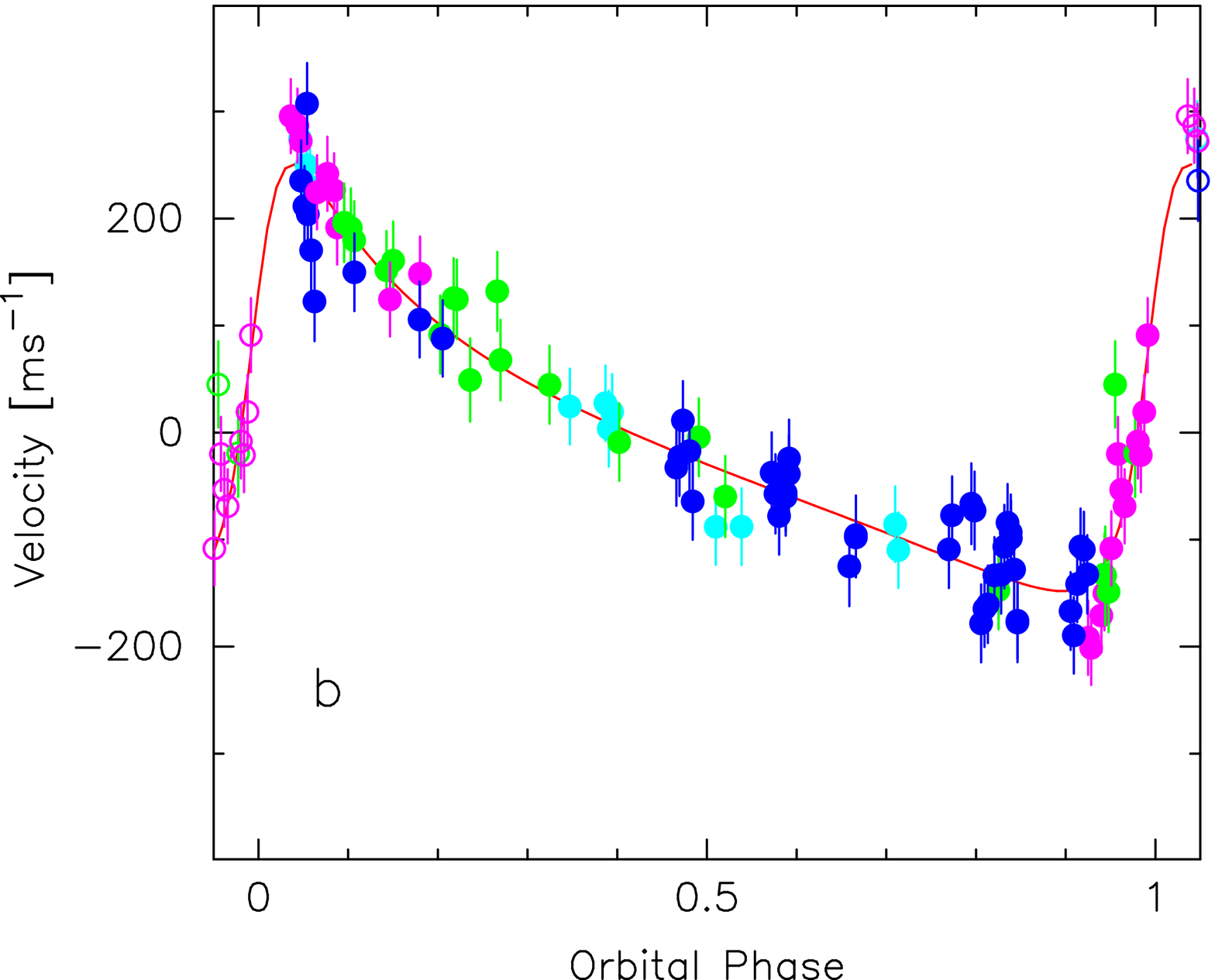}
\caption{
\textit{Top panel:} Observed RV variations and the best-fit orbital solution of HD 1666
with its residuals to the Best-fit-model.
Error bars are drawn as the summations of quadrature of observation uncertainties and assumed stellar jitter of $\sigma_{\rm jitter}=34.6\ \mps$.
The RV offsets of $-7.4\ \mps$ and 
$54.3\ \mps$ were added to HDS and HIDES data, and HIRES data, respectively.
The color legends are the same as those of Figure \ref{fig_fit1605}.
\textit{Bottom panel:} Phase-folded RV variations of HD 1666.
\label{fig_fit1666}
}
\end{figure}

The observed RVs showed complex patterns, as seen in Figure \ref{fig_fit1666}.
We applied a Lomb-Scargle periodogram to the measured RVs of HD 1666 for the determination of the initial guess of orbital parameters and the result is shown in Figure \ref{fig_prd1666}.
We found a $\sim270$ days periodicity with a confidence level higher than 99.9\% by calculating the False-Alarm-Probability (FAP) defined by 
\cite{2008MNRAS.385.1279B}.
In order to derive the Keplerian orbital motion for HD 1666, we adopted the AMOEBA method as described in the previous section.
This star should be chromospherically inactive because of the small activity-index value of $\bar{S}_{HK} = 0.15$ \citep{2010ApJ...725..875I}
and the low-strength of the cores of Ca II H K lines (Figure \ref{fig_ca_h}).
According to an empirical equation of stellar jitters, as a function of $S_{HK}$ given by 
\cite{2010ApJ...725..875I},
we estimated the RV jitter of HD 1666 to be $2.4\ \mps$.
However, as seen in Figure \ref{fig_fit1666}, the RV-residuals of the orbital fitting remain still large variation in rms of $> 30\ \mps$, compared to the typical observational uncertainty of $12\ \mps$.
In general, stellar RV variations are caused by both the Keplerian motion and stellar jitter.
The 104 data from the \hip photometry 
\citep{1997ESASP1200.....P} 
show a photometric variability of $\sigma \sim 13\ {\rm mmag}$ for an inactive HD 1666.
However, this value is comparable to the typical observational error of $15\ {\rm mmag}$, indicating that there is no significant variation photometrically for HD 1666.
This implies that there may exist additional companions that are responsible for the excess RV variations.
While a period search for the residuals to the fit detected a slightly strong signal at about ten days (bottom panel of Figure \ref{fig_prd1666}), the FAP value of $3.5\%$ does not strongly support its significance.
Indeed, although we tried an additional Keplerian-orbit fitting to the residuals with an initial guess of ten days period, neither the reduced chi-square nor the rms was improved.
More precise and additional data are required to interpret these large RV-residuals.
Then, for the present, we regarded these large residuals as the stellar jitter, $\sigma_{\rm jitter}=34.6\ \mps$, which makes the reduced chi-square unity.
Finally, we obtained the orbital period of $P=270.0^{+0.8}_{-0.9}$ days, the velocity semi-amplitude $K_1 = 199.4^{+8.8}_{5.5}\mps$, and the eccentricity $e=0.63^{+0.03}_{-0.02}$ as the best-fit orbital parameters for HD 1666 b.
With the stellar mass of $1.50\pm0.07\msun$, we found that HD 1666 b has $\msini = 6.43\mjup$ at the semi-major axis of $a_{\rm p}=0.94$ AU. 
The best-fit parameters and their errors are listed in Table \ref{tbl_prm}.

\subsection{HD 67087}
\label{hd67087}
HD 67087 (HIP39767) is a F-type main-sequence star with $V=8.05$ and a color index of $B-V=0.53$.
The \textit{Hipaarcos} parallax of 11.26 mas corresponds to a distance of 88.8 pc, i.e., the absolute V magnitude of $M_V=3.31$.
The SME analysis yields the stellar parameters of $T_{\rm eff}=6330\pm47$ K, $\vsini = 9.7\pm0.5\ \kmps$, [Fe/H]$=+0.25\pm0.04$, and $\logg=4.19\pm0.06$.
Using $T_{\rm eff}$, $B-V$, $M_V$, and $B.C.=-0.010$, we have estimated the bolometric luminosity and stellar radius to be $L=3.47\pm 0.75L_{\odot}$ and $R_{*}=1.55\pm0.32\rsun$ from the Stefan-Boltzmann law.
The stellar mass and age were estimated to be $M_{*}=1.36\pm0.04M_{\odot}$ and $1.45\pm0.51$ Gyr from stellar evolution tracks, respectively, (see also Section \ref{subsec_1605}).

We have monitored HD 67087 using HDS and HIDES for about 9 years (from December 2004 to May 2014). 
We obtained 51 RV data in total, 12 and 39 of which were observed by HDS and HIDES-F excluding poor observations, respectively.
The observation dates, RVs, internal errors, and observatory sites are listed in Table \ref{tbl_obs67087}.
The observed RVs show a clear significant periodicity in addition to that of about one-year in Figure \ref{fig_fit67087}.
Assuming the stellar jitter of $4\ \mps$ based on the low emission strength in the core of Ca II HK lines (Figure \ref{fig_ca_h}), we fitted two-Keplerian orbits to the measured RVs.
However, we cannot obtain the RV semi-amplitude and the minimum mass of a planet, via the bootstrap method due to lack of RV peaks for the outer orbit.
Hence, we applied the MCMC procedure following
\cite{2013PASJ...65...85S}
to the RVs and determined two-Keplerian orbits from the posterior distribution functions.

We considered the medians at the $1\sigma$ confidence level as the best-fit values.
The rms to the fitting was $11.8\ \mps$, and the square root of reduced $\chi^2$ of $\sqrt{\chi^2_\nu}=0.91$, indicating that our orbital solutions are preferable within observational errors.
With the stellar mass of $M_*=1.36\pm0.04\msun$, $P=352.27^{+1.7}_{-1.6}$, and $K_1 = 73.6^{+4.4}_{-3.9} \mps$, we found $a_{\rm p}=1.08$ AU and $\msini = 3.06\mjup$ for the inner planet HD 67087 b.
The outer planet HD 67087 c has the orbital period $P=2374^{+193}_{-156}$ days in an eccentric orbit of $e = 0.76^{+0.17}_{-0.24}$ and $K_1 = 93.3^{+151.4}_{-40.8}\ \mps$, and hence, $\msini = 4.85^{+10.0}_{-3.61}\mjup$ at $a_{\rm p}=3.86^{+0.43}_{-0.37}$ AU.
Figure \ref{fig_fit67087} shows observations and best-fit orbital solutions.
The orbital solutions and their errors are summarized in Table \ref{tbl_prm}.

Although the phase-folded RV variation of HD 67087 c appears a simple linear RV trend, the actual time variation in RVs repeats this trend more than one within the duration of observations.
We tried the Keplerian fitting with one-planet + linear trend model and obtained a reduced chi-square of $\chi_{\nu}^2 = 3.4$.
Therefore, the two-planet model may be preferred for this system, considering the reduced chi-square of $\chi_{\nu}^2 = 0.81$.

\section{Discussion}
\label{sec_disc}
In this paper, we have reported five newly-discovered Jovian planets around metal-rich stars from our long-term planet-search at the Subaru telescope, the OAO 188 cm telescope and the Keck telescope.
Two multi-Jovian-planet systems of HD 1605 and HD 67087, and the eccentric massive planetary system of HD 1666 have been identified.
All of the host stars have high metallicity ([Fe/H]$> +0.2$) and relatively large masses of  $M_* > 1.3 \msun$.

HD 1605 is a subgiant star and its mass is estimated to be $1.3 \msun$.
We have confirmed the two Jovian-mass planets on the hierarchical system with a linear RV trend.
According to the SIMBAD database, no stellar companions has been reported within 1.1 arcsec ($< 100\ {\rm AU}$) from HD 1605.
The linear trend of RV that we found suggests the presence of an unknown and maybe low-mass companion.
Given that the acceleration of RVs is constant, we can give an order-of-magnitude relation between the linear trend and the properties of the outer companion
\citep{2009ApJ...703L..99W},
\begin{equation}
\frac{M_c\sin i_c}{a_c} \sim \frac{\dot \gamma}{G} = (0.028^{+0.001}_{-0.001}) M_J\ {\rm AU}^{-2},
\end{equation}
where $M_c$ is the companion mass, $i_c$ is the orbital inclination relative to the line of sight, $a_c$ is the orbital distance, and $\dot \gamma$ is the acceleration of RVs.
The HD 1605 system is also a rare planetary system in terms of their low-eccentricity orbits.
There are 44 systems which hold multiple Jovian-planets with masses of $\msini > 0.3\ \mjup$ 
\citep[the Exoplanets Orbit Database: ][]{2011PASP..123..412W},
and only four of them are known to have multiple Jovian planets with low-eccentricities ($e<0.1$) in the absence of hot Jupiters 
\citep[47 UMa, 55 Cnc, $\mu$ Ara and HD 159243:][respectively]
{2010MNRAS.403..731G,2012ApJ...759...19E,2007A&A...462..769P,2014A&A...563A..22M}.

The two-planet system of HD 67087 is also a hierarchical system.
HD 67087 has the mass of $1.36 \msun$ and metallicity of [Fe/H] = $+0.25$.
We have found two super-Jupiter-mass planets at $1.08$ AU and $3.86$ AU around HD 67087.
Interestingly, HD 67087 c may be one of the most eccentric Jovian-planets among those with $a_{\rm p} > 3$ AU and also among multiple Jovian-planet-systems, in contrast to HD 1605 c (Figure \ref{fig_a_e}).
Although Jovian planets similar to HD 67087 c are seen in other systems such as HD 181433 d 
\citep{2009A&A...496..527B} 
and HD 217107 c 
\citep{2005ApJ...632..638V,2009ApJ...693.1084W},
hot-Jupiters were found in their inner orbits.
Thus, the origin of those eccentric outer planets can be explained by planet-planet interactions.
On the other hand, in the case of HD 67087 which has no hot-Jupiter, the existence of an eccentric outer planet around HD 67087 may be interpreted by the excitation of its eccentricity due to secular perturbations from an unseen substellar companion in distant orbit, the so-called Kozai mechanism
\citep{1962AJ.....67..591K}.
Although the lack of observations does not lead to the accurate determination of the orbit of HD 67087 c, the two planets in this system have mutually crossing orbits at present.
Hence, the dynamical stability of HD 67087 system should be investigated with additional RVs in the future work. 
This allows us to constrain the mass and the eccentricity of HD 67087 c more strictly.
Although we did not find any long-term RV trend due to an additional companion from the residuals to our best-fit model, a relatively-young HD 67087 still becomes a promising target for direct-imaging surveys.
If the existence of an outer companion is confirmed, HD 67087 c should be a strong evidence of orbital evolution via Kozai mechanism.

The F7 dwarf HD 1666 has a relatively-large mass of $1.5\msun$ and a very high metallicity of $\feh=+0.37$.
HD 1666 hosts a massive, eccentric planet HD 1666 b.
One of the most remarkable features of HD 1666 system is the mass of a host star.
It is generally known that early-type main-sequence stars which have masses higher than that of the Sun such as A-dwarfs (typically $1.5\msun$--$2\msun$) are not suitable for RV measurements because of Doppler-broadening of absorption lines due to their rapid rotation and the deficiency of absorption lines due to high effective temperatures.
RV surveys rarely detected short-period planets around evolved high-mass stars
\citep[e.g.][]{2010ApJ...721L.153J}.
However, HD 1666 has a cooler effective temperature and lower rotational velocity.
A high metallicity in the stellar atmosphere causes a larger stellar radius at a given luminosity and stellar mass.
A metal-rich star like HD 1666 with $\feh = +0.37$ tends to have a lower effective temperature 
\citep[e.g.][]{2000A&AS..141..371G,2001ApJS..136..417Y}.
As seen in Figure 17 of \cite{2005ApJS..159..141V}, $T_{\rm eff}$ is decreased by about 500K, when $\feh$ increases by 0.5 dex.
Our target selection for a metal-rich star based on the color-index (typically $B-V$) leads to selecting a high-mass star, and vice versa.
Figure \ref{fig_m_m} shows the planetary mass distribution as a function of stellar mass.
All of the Jovian planets in this figure show those around main sequence stars ($\log g > 4.0$) detected by the RV method. 
Figure \ref{fig_m_m} indicates that HD 1666 is one of the most massive main-sequence stars harboring planetary companions.
Therefore, metal-rich and massive main-sequence stars such as HD 1666 can be suitable targets not only to unveil a correlation between planetary properties and the stellar metallicity, but also to provide information about the dependence of planet population on stellar mass.

\begin{figure}
\epsscale{.80}
\plotone{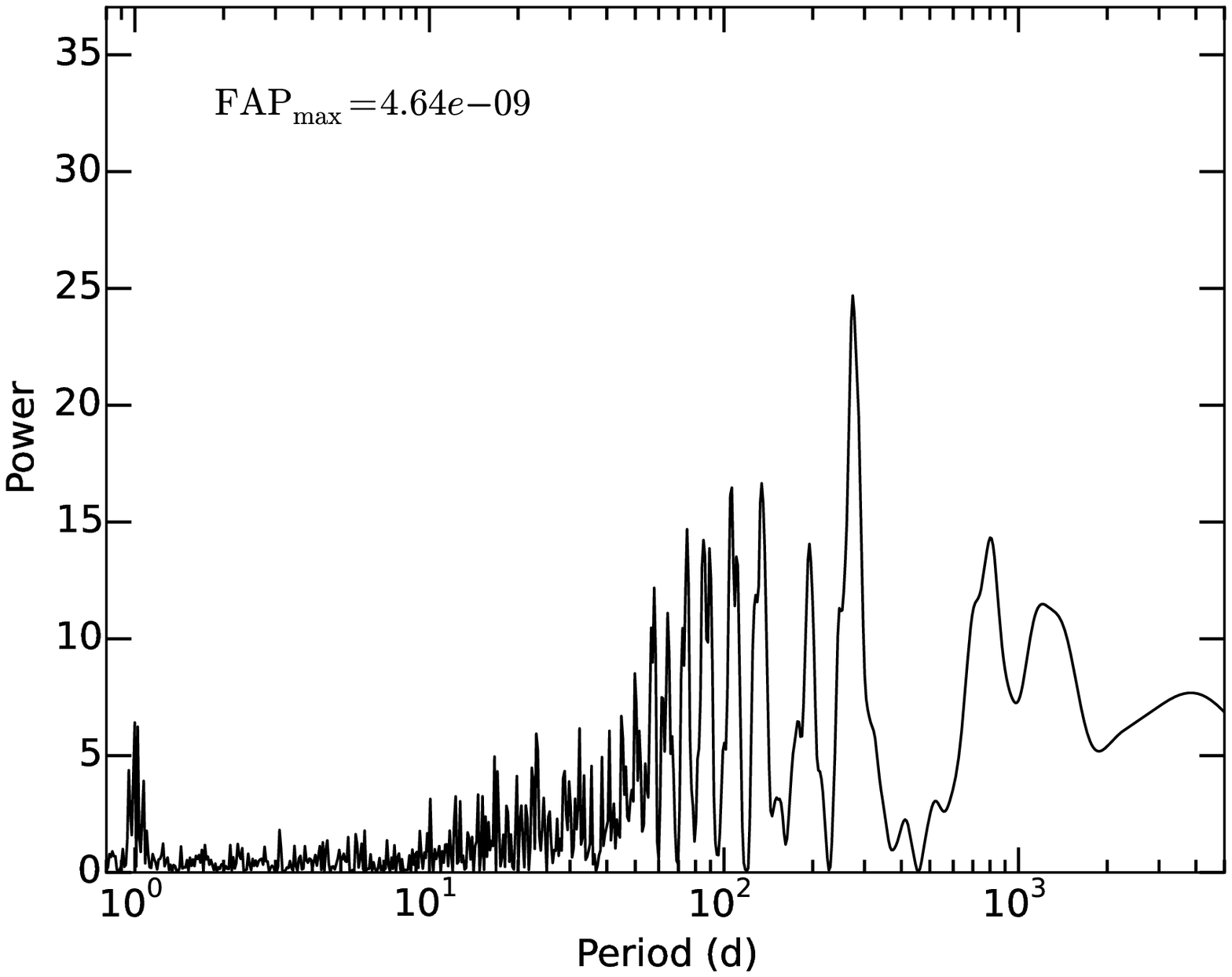}
\plotone{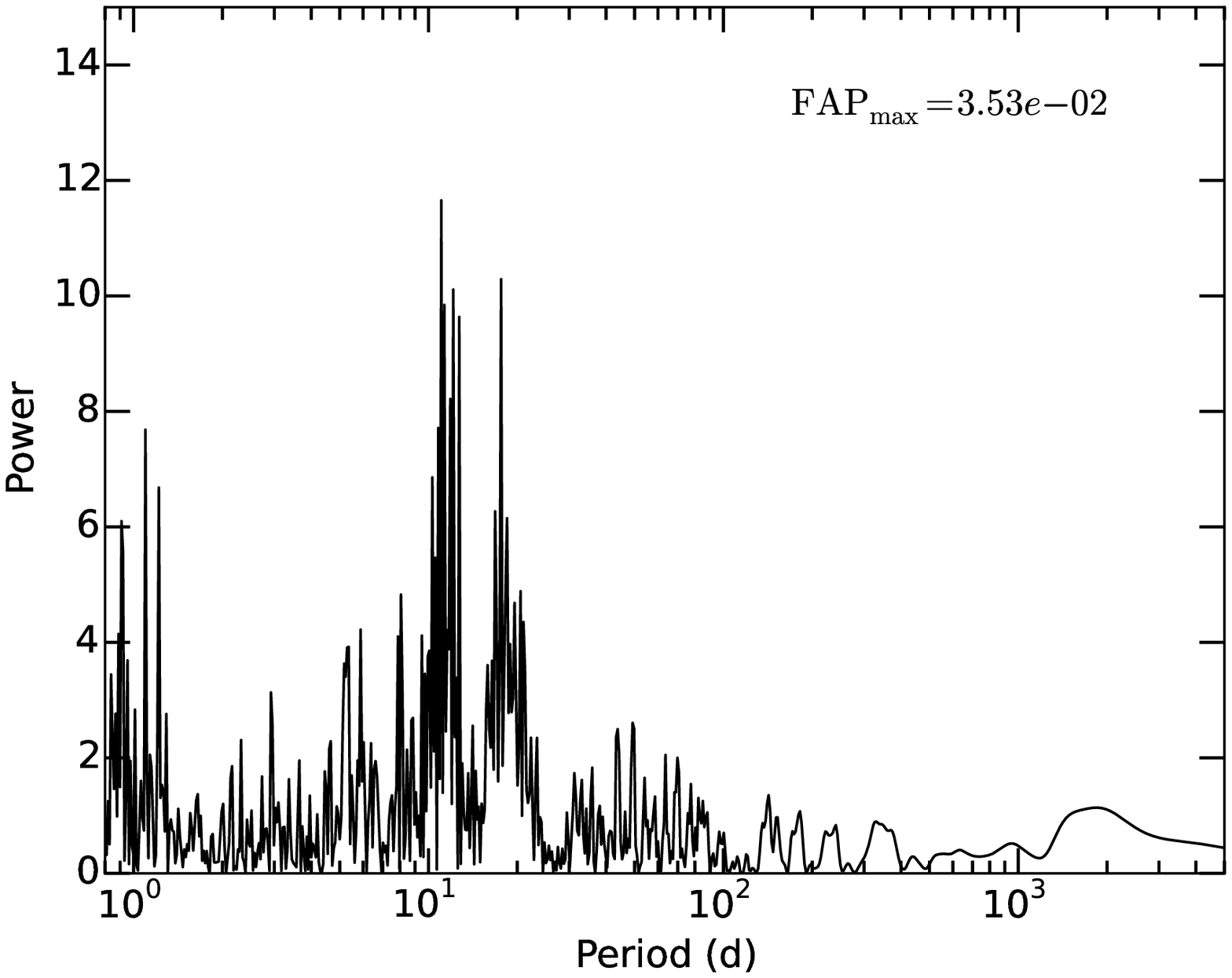}
\caption{
\textit{Upper panel:} the Lomb-Scargle periodogram for RV measurements on HD 1666.
FAP on the maximum peak ($\sim 270$ days) was calculated based on the definition from \cite{2008MNRAS.385.1279B}.
\textit{Lower panel:} same for the residuals to the Keplerian fit.
\label{fig_prd1666}
}
\end{figure}

\begin{figure}
\epsscale{.65}
\plotone{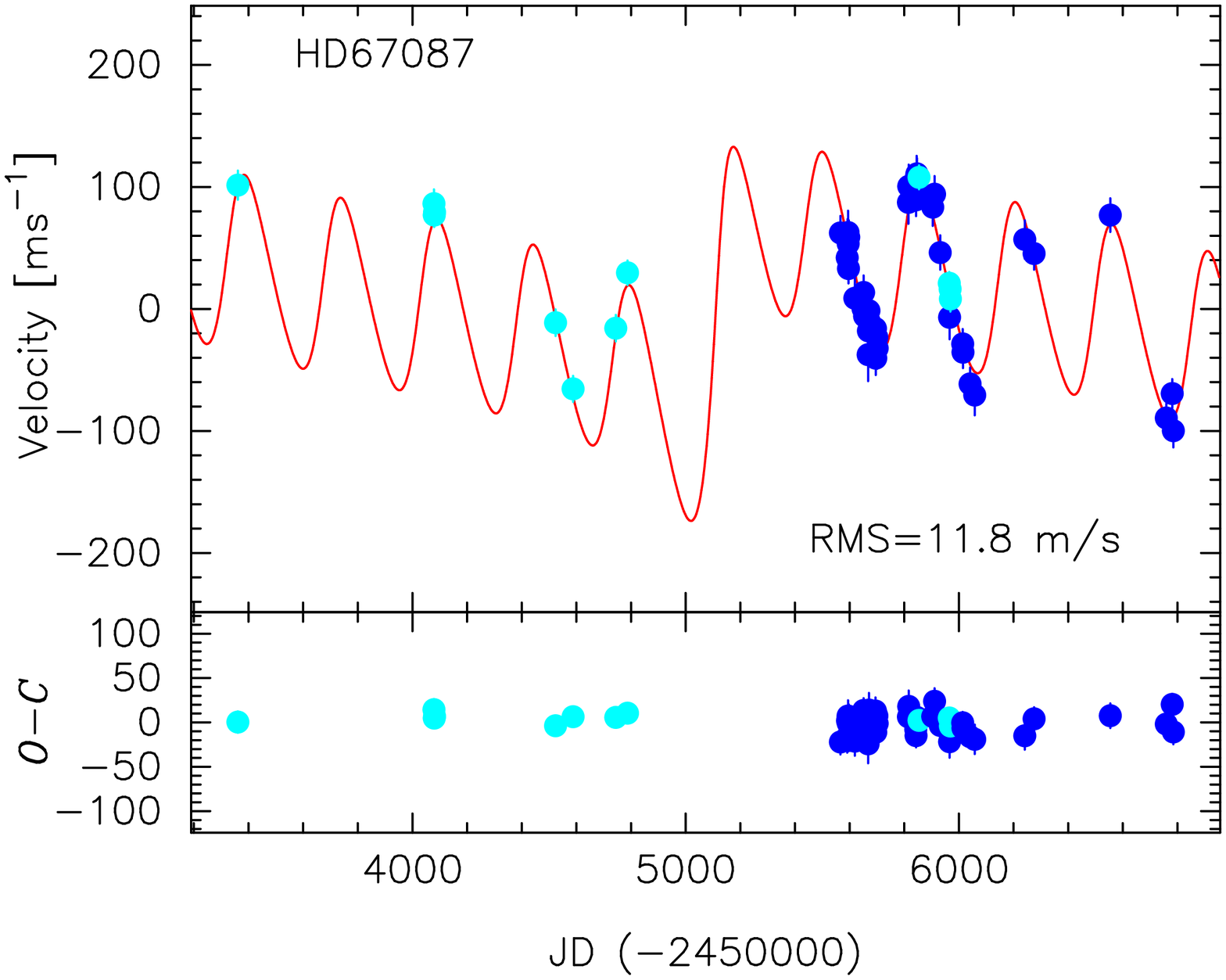}
\plotone{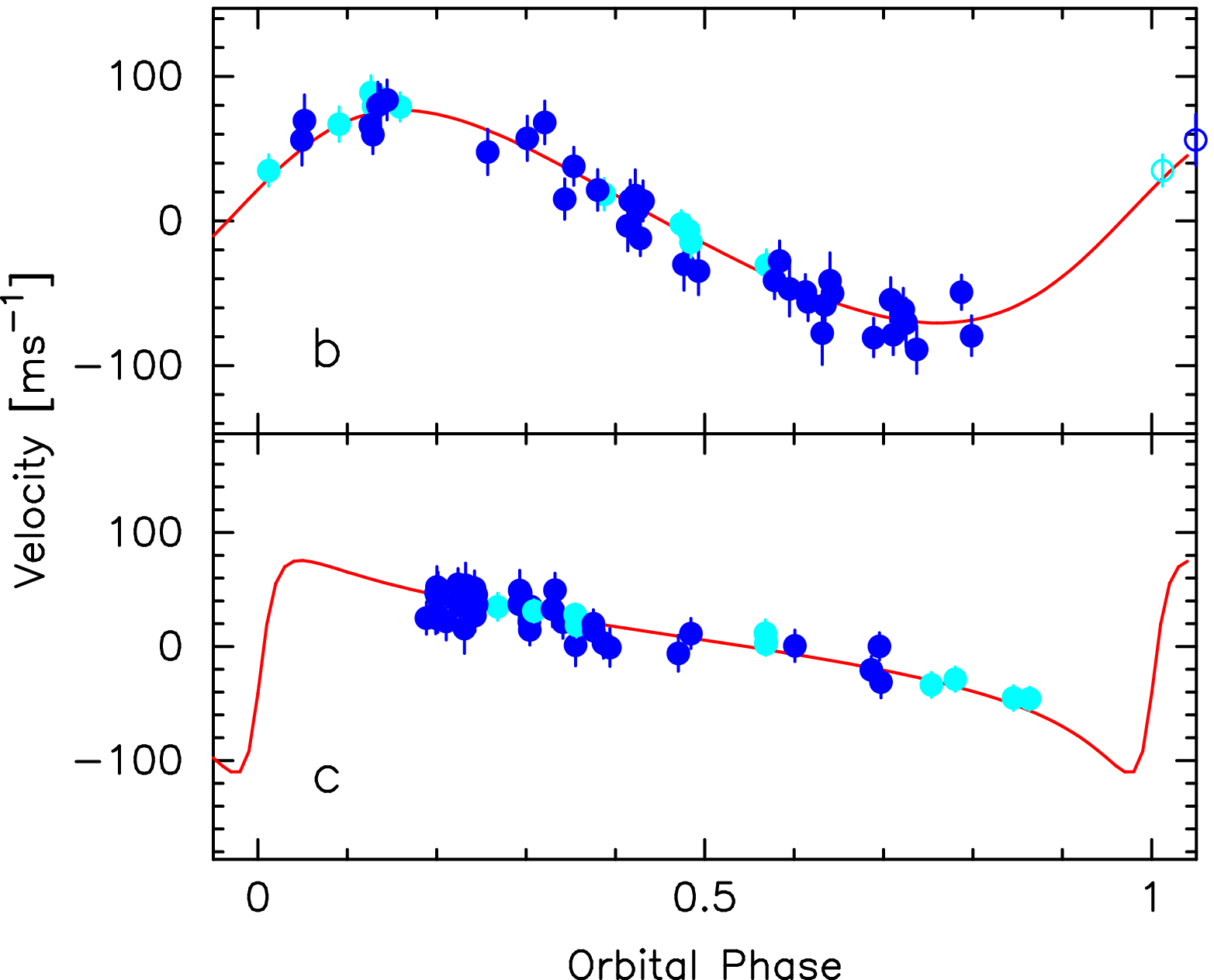}
\caption{
\textit{Top panel:} Observed RV variations and the best-fit orbital solution of HD 67087 with its residuals to the best-fit model.
The color legends are the same as Figure \ref{fig_fit1605}.
\textit{Bottom panel:} Phased RV variations of HD 67087.
\label{fig_fit67087}
}
\end{figure}

\begin{figure}
\epsscale{.80}
\plotone{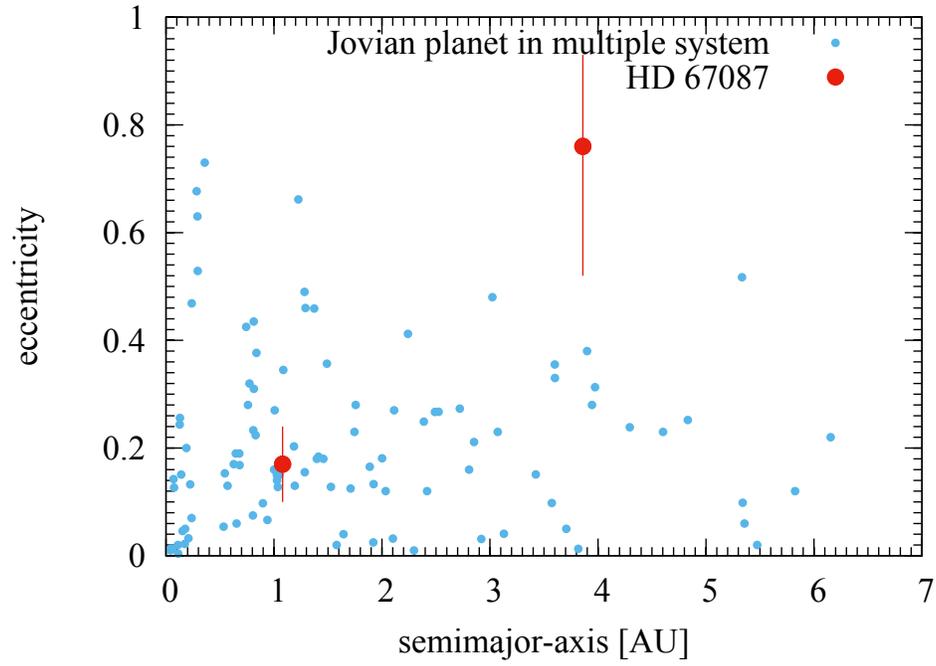}
\caption{Planet distribution in $a_{\rm p}$-$e$ diagram.
The planets with $\msini > 0.1 \mjup$ in multiple systems are shown.
Only the error bars of HD 67087 planets are drawn for clarity.
---http://exoplanets.org
\label{fig_a_e}
}
\end{figure}

\begin{figure}
\epsscale{.80}
\plotone{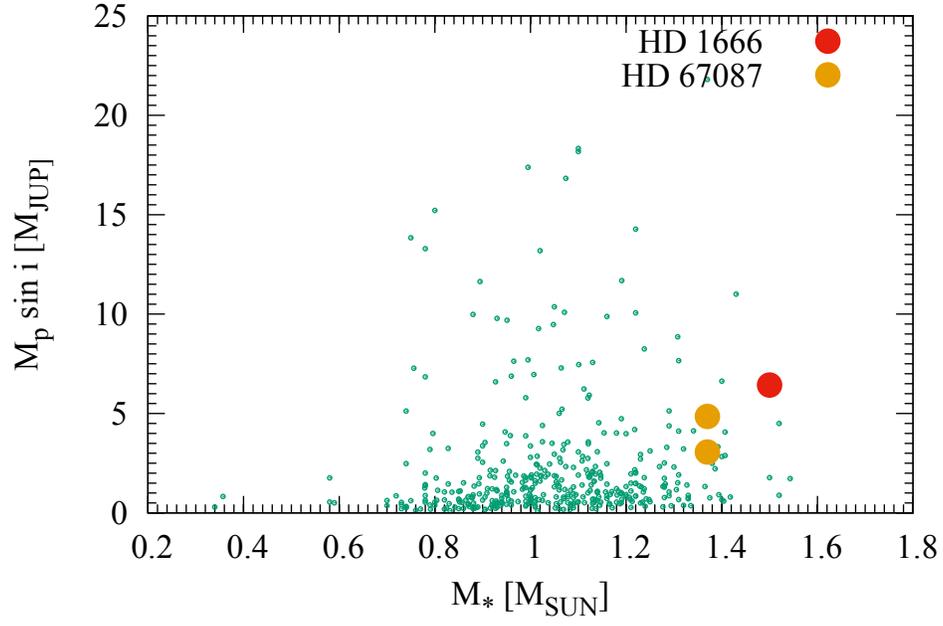}
\caption{Distribution of minimum masses of exoplanets and masses of their host stars.
Only Jovian planets around main-sequence stars ($\msini > 0.1 \mjup$, $\log g > 4.0$) are plotted 
---http://exoplanets.org.
Planets around HD 1666 and HD 67087 are shown with {\it filled red circle} 
and {\it filled orange circle}, respectively.
\label{fig_m_m}
}
\end{figure}

\begin{figure}
\epsscale{.80}
\plotone{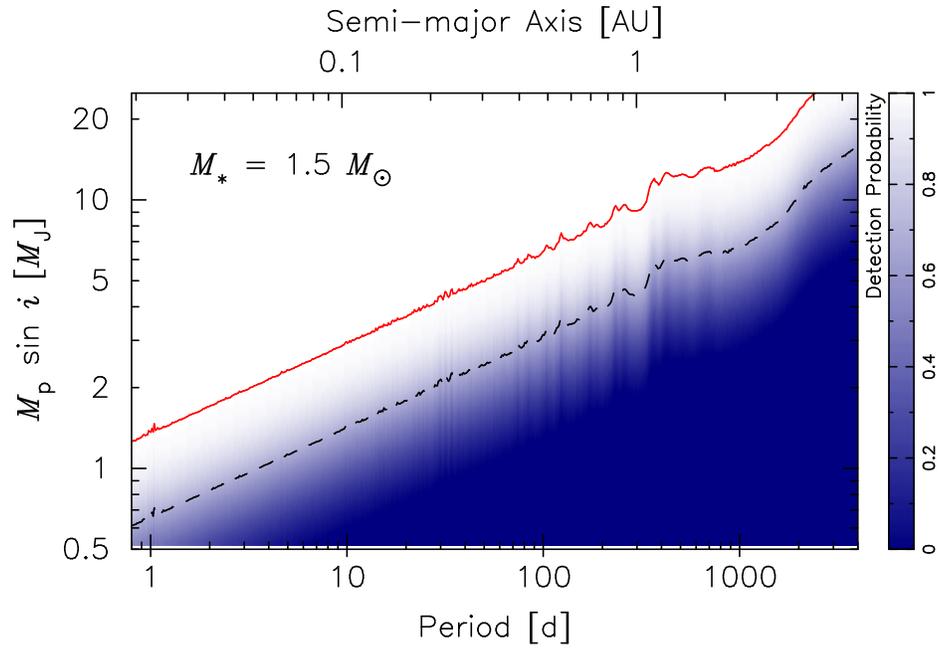}
\caption
{
\label{fig:dl_1666}
Detection probability map calculated from the RV residuals of HD 1666. 
The \textit{red solid line} and the \textit{black dashed line} represent $3\-\sigma$(99.73\%) and $1\-\sigma$(68.27\%) detection probability, respectively.
}
\end{figure}

Another important feature of HD 1666 system is the eccentric planet of HD 1666 b. 
As mentioned before, such an eccentric planet can be interpreted as a by-product of gravitational interactions among planets and/or stellar companions, although we have not yet found any inner or outer companion.
The detection-limit analysis following \cite{2009A&A...495..335L} is shown at Figure \ref{fig:dl_1666}.
We here, briefly describe about this analysis method:
(1) for each (mass; period) grid, generate 1000 synthetic RV-data-sets along the actual observation times assuming circular orbit and different times of passage at periastron;
(2) add a random noise corresponding real observational errors;
(3) compare the averaged rms of virtual data-sets and that of real residual data.
If the rms of the real residuals is lower than the average rms of virtual data sets, the given planet can be detected.
The detection probability can be calculated as a level of rms excess compared to that of the real data.
Figure \ref{fig:dl_1666} shows that while the stellar jitter ($\sigma_{\rm jitter}=35\ \mps$) may prevent us from detecting low-mass planets, it cannot be an obstacle for the detection of a hot-Jupiter with the mass of $\msini > 3\mjup$ with $P<10$ days at a $3\sigma$ confidence level.
As seen in Figure \ref{fig_prd1666}, the non-detection for short periodicity from the residuals indicates that (i) if HD 1666 b experienced planet-planet scattering, an scattered planet is likely to be outside HD 1666 b or (ii) the eccentricity of HD 1666 b was pumped up by an outer companion via the Kozai mechanism.
Indeed, the RV amplitude induced by a distant substellar companion can be smaller than this jitter.
The future direct-imaging survey or correction for stellar jitters with careful analyses of activity-induced RVs 
\citep[e.g.][]{2011MNRAS.411.1953P,2012A&A...545A.109B,
2013ApJ...770..133H,2014A&A...566A..35S} 
helps us verify the existence of the additional companion(s).

\acknowledgments
We acknowledge the referee for fruitful comments and suggestions.
This research is based on data collected at the Okayama Astrophysical Observatory and the Subaru telescope which is operated by National Astronomical Observatory of Japan, and at the Keck telescope which is operated as a scientific partner ship  among the California Institute of Technology, the University of California and the National Aeronautics and Space Administration.
We are grateful to all the staff members of OAO for their support during the observations.
We are grateful to Akito Tajitsu for his expertise and support of HDS observations at the Subaru telescope.
We also gratefully acknowledge the efforts and dedication of all the Keck Observatory staffs.
This research has made use of the SIMBAD database, operated at CDS, Strasbourg, France, the Exoplanet Data Explorer at exoplanets.org, and The Extrasolar Planets Encyclopedia (exoplanet.eu).
H. H. is supported by Grant-in-Aid for Japan Society for the Promotion of Science (JSPS) Fellows (No. 20115548) from the Ministry of Education, Culture, Sports, Science and Technology (MEXT) of Japan and also supported by the Global COE program ``From Earth to Earths''.
Y. H. is supported by Grant-in-Aid for JSPS Fellows (No. 25000465) from MEXT and Grant-in-Aid for Scientific Research on Innovative Areas (No. 26103711) from MEXT.
E. K. is supported by Grants-in-Aid for Scientific Research (No. 20540240) from the Japan Society for the Promotion of Science (JSPS).
H. I. is supported by Grant-In-Aid for Scientific Research (A) 23244038 from JSPS.
The authors wish to acknowledge people in Hawaii for their hospitality and native Hawaiian ancestry on whose sacred mountain of Mauna Kea we are privileged to be visitors.

{\it Facilities:} \facility{OAO (HIDES)}, \facility{Subaru (HDS)}, \facility{Keck (HIRES)}.

\bibliography{library}

\begin{thebibliography}{67}
\expandafter\ifx\csname natexlab\endcsname\relax\def\natexlab#1{#1}\fi

\bibitem[{{Adibekyan} {et~al.}(2013){Adibekyan}, {Figueira}, {Santos},
  {Mortier}, {Mordasini}, {Delgado Mena}, {Sousa}, {Correia}, {Israelian}, \&
  {Oshagh}}]{2013A&A...560A..51A}
{Adibekyan}, V.~Z., {Figueira}, P., {Santos}, N.~C., {et~al.} 2013, \aap, 560,
  A51

\bibitem[{{Baluev}(2008)}]{2008MNRAS.385.1279B}
{Baluev}, R.~V. 2008, \mnras, 385, 1279

\bibitem[{{Boisse} {et~al.}(2012){Boisse}, {Bonfils}, \&
  {Santos}}]{2012A&A...545A.109B}
{Boisse}, I., {Bonfils}, X., \& {Santos}, N.~C. 2012, \aap, 545, A109

\bibitem[{{Bouchy} {et~al.}(2009){Bouchy}, {Mayor}, {Lovis}, {Udry}, {Benz},
  {Bertaux}, {Delfosse}, {Mordasini}, {Pepe}, {Queloz}, \&
  {Segransan}}]{2009A&A...496..527B}
{Bouchy}, F., {Mayor}, M., {Lovis}, C., {et~al.} 2009, \aap, 496, 527

\bibitem[{{Bowler} {et~al.}(2010){Bowler}, {Johnson}, {Marcy}, {Henry}, {Peek},
  {Fischer}, {Clubb}, {Liu}, {Reffert}, {Schwab}, \&
  {Lowe}}]{2010ApJ...709..396B}
{Bowler}, B.~P., {Johnson}, J.~A., {Marcy}, G.~W., {et~al.} 2010, \apj, 709,
  396

\bibitem[{{Butler} {et~al.}(1996){Butler}, {Marcy}, {Williams}, {McCarthy},
  {Dosanjh}, \& {Vogt}}]{1996PASP..108..500B}
{Butler}, R.~P., {Marcy}, G.~W., {Williams}, E., {et~al.} 1996, \pasp, 108, 500

\bibitem[{{Cumming} {et~al.}(2008){Cumming}, {Butler}, {Marcy}, {Vogt},
  {Wright}, \& {Fischer}}]{2008PASP..120..531C}
{Cumming}, A., {Butler}, R.~P., {Marcy}, G.~W., {et~al.} 2008, \pasp, 120, 531

\bibitem[{{Cumming} {et~al.}(1999){Cumming}, {Marcy}, \&
  {Butler}}]{1999ApJ...526..890C}
{Cumming}, A., {Marcy}, G.~W., \& {Butler}, R.~P. 1999, \apj, 526, 890

\bibitem[{{Endl} {et~al.}(2012){Endl}, {Robertson}, {Cochran}, {MacQueen},
  {Brugamyer}, {Caldwell}, {Wittenmyer}, {Barnes}, \&
  {Gullikson}}]{2012ApJ...759...19E}
{Endl}, M., {Robertson}, P., {Cochran}, W.~D., {et~al.} 2012, \apj, 759, 19

\bibitem[{{Ercolano} \& {Clarke}(2010)}]{2010MNRAS.402.2735E}
{Ercolano}, B., \& {Clarke}, C.~J. 2010, \mnras, 402, 2735

\bibitem[{{Fischer} \& {Valenti}(2005)}]{2005ApJ...622.1102F}
{Fischer}, D.~A., \& {Valenti}, J. 2005, \apj, 622, 1102

\bibitem[{{Fischer} {et~al.}(2005){Fischer}, {Laughlin}, {Butler}, {Marcy},
  {Johnson}, {Henry}, {Valenti}, {Vogt}, {Ammons}, {Robinson}, {Spear},
  {Strader}, {Driscoll}, {Fuller}, {Johnson}, {Manrao}, {McCarthy},
  {Mu{\~n}oz}, {Tah}, {Wright}, {Ida}, {Sato}, {Toyota}, \&
  {Minniti}}]{2005ApJ...620..481F}
{Fischer}, D.~A., {Laughlin}, G., {Butler}, P., {et~al.} 2005, \apj, 620, 481

\bibitem[{{Fischer} {et~al.}(2007){Fischer}, {Vogt}, {Marcy}, {Butler}, {Sato},
  {Henry}, {Robinson}, {Laughlin}, {Ida}, {Toyota}, {Omiya}, {Driscoll},
  {Takeda}, {Wright}, \& {Johnson}}]{2007ApJ...669.1336F}
{Fischer}, D.~A., {Vogt}, S.~S., {Marcy}, G.~W., {et~al.} 2007, \apj, 669, 1336

\bibitem[{{Flower}(1996)}]{1996ApJ...469..355F}
{Flower}, P.~J. 1996, \apj, 469, 355

\bibitem[{{Giguere} {et~al.}(2012){Giguere}, {Fischer}, {Howard}, {Johnson},
  {Henry}, {Wright}, {Marcy}, {Isaacson}, {Hou}, \&
  {Spronck}}]{2012ApJ...744....4G}
{Giguere}, M.~J., {Fischer}, D.~A., {Howard}, A.~W., {et~al.} 2012, \apj, 744,
  4

\bibitem[{{Girardi} {et~al.}(2002){Girardi}, {Bertelli}, {Bressan}, {Chiosi},
  {Groenewegen}, {Marigo}, {Salasnich}, \& {Weiss}}]{2002A&A...391..195G}
{Girardi}, L., {Bertelli}, G., {Bressan}, A., {et~al.} 2002, \aap, 391, 195

\bibitem[{{Girardi} {et~al.}(2000){Girardi}, {Bressan}, {Bertelli}, \&
  {Chiosi}}]{2000A&AS..141..371G}
{Girardi}, L., {Bressan}, A., {Bertelli}, G., \& {Chiosi}, C. 2000, \aaps, 141,
  371

\bibitem[{{Gonzalez}(1997)}]{1997MNRAS.285..403G}
{Gonzalez}, G. 1997, \mnras, 285, 403

\bibitem[{{Gregory} \& {Fischer}(2010)}]{2010MNRAS.403..731G}
{Gregory}, P.~C., \& {Fischer}, D.~A. 2010, \mnras, 403, 731

\bibitem[{{Harakawa} {et~al.}(2010){Harakawa}, {Sato}, {Fischer}, {Ida},
  {Omiya}, {Johnson}, {Marcy}, {Toyota}, {Hori}, \&
  {Howard}}]{2010ApJ...715..550H}
{Harakawa}, H., {Sato}, B., {Fischer}, D.~A., {et~al.} 2010, \apj, 715, 550

\bibitem[{{Hatzes}(2013)}]{2013ApJ...770..133H}
{Hatzes}, A.~P. 2013, \apj, 770, 133

\bibitem[{{Howard} {et~al.}(2010){Howard}, {Marcy}, {Johnson}, {Fischer},
  {Wright}, {Isaacson}, {Valenti}, {Anderson}, {Lin}, \&
  {Ida}}]{2010Sci...330..653H}
{Howard}, A.~W., {Marcy}, G.~W., {Johnson}, J.~A., {et~al.} 2010, Science, 330,
  653

\bibitem[{{Ida} \& {Lin}(2004)}]{2004ApJ...604..388I}
{Ida}, S., \& {Lin}, D.~N.~C. 2004, \apj, 604, 388

\bibitem[{{Ida} {et~al.}(2013){Ida}, {Lin}, \&
  {Nagasawa}}]{2013ApJ...775...42I}
{Ida}, S., {Lin}, D.~N.~C., \& {Nagasawa}, M. 2013, \apj, 775, 42

\bibitem[{{Isaacson} \& {Fischer}(2010)}]{2010ApJ...725..875I}
{Isaacson}, H., \& {Fischer}, D. 2010, \apj, 725, 875

\bibitem[{{Izumiura}(1999)}]{1999oaaf.conf...77I}
{Izumiura}, H. 1999, in Observational Astrophysics in Asia and its Future, ed.
  P.~S. {Chen}, 77

\bibitem[{{Johnson} {et~al.}(2010{\natexlab{a}}){Johnson}, {Aller}, {Howard},
  \& {Crepp}}]{2010PASP..122..905J}
{Johnson}, J.~A., {Aller}, K.~M., {Howard}, A.~W., \& {Crepp}, J.~R.
  2010{\natexlab{a}}, \pasp, 122, 905

\bibitem[{{Johnson} {et~al.}(2010{\natexlab{b}}){Johnson}, {Bowler}, {Howard},
  {Henry}, {Marcy}, {Isaacson}, {Brewer}, {Fischer}, {Morton}, \&
  {Crepp}}]{2010ApJ...721L.153J}
{Johnson}, J.~A., {Bowler}, B.~P., {Howard}, A.~W., {et~al.}
  2010{\natexlab{b}}, \apjl, 721, L153

\bibitem[{{Kambe} {et~al.}(2002){Kambe}, {Sato}, {Takeda}, {Ando}, {Noguchi},
  {Aoki}, {Izumiura}, {Wada}, {Masuda}, {Okada}, {Shimizu}, {Watanabe},
  {Yoshida}, {Honda}, \& {Kawanomoto}}]{2002PASJ...54..865K}
{Kambe}, E., {Sato}, B., {Takeda}, Y., {et~al.} 2002, \pasj, 54, 865

\bibitem[{{Kambe} {et~al.}(2013){Kambe}, {Yoshida}, {Izumiura}, {Koyano},
  {Nagayama}, {Shimizu}, {Okada}, {Okita}, {Sakamoto}, {Sato}, \&
  {Yamamuro}}]{2013PASJ...65...15K}
{Kambe}, E., {Yoshida}, M., {Izumiura}, H., {et~al.} 2013, \pasj, 65, 15

\bibitem[{{Kozai}(1962)}]{1962AJ.....67..591K}
{Kozai}, Y. 1962, \aj, 67, 591

\bibitem[{{Lagrange} {et~al.}(2009){Lagrange}, {Desort}, {Galland}, {Udry}, \&
  {Mayor}}]{2009A&A...495..335L}
{Lagrange}, A.-M., {Desort}, M., {Galland}, F., {Udry}, S., \& {Mayor}, M.
  2009, \aap, 495, 335

\bibitem[{{Lomb}(1976)}]{1976Ap&SS..39..447L}
{Lomb}, N.~R. 1976, \apss, 39, 447

\bibitem[{{L{\'o}pez-Morales} {et~al.}(2008){L{\'o}pez-Morales}, {Butler},
  {Fischer}, {Minniti}, {Shectman}, {Takeda}, {Adams}, {Wright}, \&
  {Arriagada}}]{2008AJ....136.1901L}
{L{\'o}pez-Morales}, M., {Butler}, R.~P., {Fischer}, D.~A., {et~al.} 2008, \aj,
  136, 1901

\bibitem[{{Marcy} \& {Butler}(1992)}]{1992PASP..104..270M}
{Marcy}, G.~W., \& {Butler}, R.~P. 1992, \pasp, 104, 270

\bibitem[{{Mayor} {et~al.}(2011){Mayor}, {Marmier}, {Lovis}, {Udry},
  {S{\'e}gransan}, {Pepe}, {Benz}, {Bertaux}, {Bouchy}, {Dumusque}, {Lo Curto},
  {Mordasini}, {Queloz}, \& {Santos}}]{2011arXiv1109.2497M}
{Mayor}, M., {Marmier}, M., {Lovis}, C., {et~al.} 2011, ArXiv e-prints

\bibitem[{{Mordasini} {et~al.}(2009){Mordasini}, {Alibert}, \&
  {Benz}}]{2009A&A...501.1139M}
{Mordasini}, C., {Alibert}, Y., \& {Benz}, W. 2009, \aap, 501, 1139

\bibitem[{{Mortier} {et~al.}(2012){Mortier}, {Santos}, {Sozzetti}, {Mayor},
  {Latham}, {Bonfils}, \& {Udry}}]{2012A&A...543A..45M}
{Mortier}, A., {Santos}, N.~C., {Sozzetti}, A., {et~al.} 2012, \aap, 543, A45

\bibitem[{{Moutou} {et~al.}(2014){Moutou}, {H{\'e}brard}, {Bouchy}, {Arnold},
  {Santos}, {Astudillo-Defru}, {Boisse}, {Bonfils}, {Borgniet}, {Delfosse},
  {D{\'{\i}}az}, {Ehrenreich}, {Forveille}, {Gregorio}, {Labrevoir},
  {Lagrange}, {Montagnier}, {Montalto}, {Pepe}, {Sahlmann}, {Santerne},
  {S{\'e}gransan}, {Udry}, \& {Vanhuysse}}]{2014A&A...563A..22M}
{Moutou}, C., {H{\'e}brard}, G., {Bouchy}, F., {et~al.} 2014, \aap, 563, A22

\bibitem[{{Nagasawa} \& {Ida}(2011)}]{2011ApJ...742...72N}
{Nagasawa}, M., \& {Ida}, S. 2011, \apj, 742, 72

\bibitem[{{Nagasawa} {et~al.}(2008){Nagasawa}, {Ida}, \&
  {Bessho}}]{2008ApJ...678..498N}
{Nagasawa}, M., {Ida}, S., \& {Bessho}, T. 2008, \apj, 678, 498

\bibitem[{Nelder \& Mead(1965)}]{Nelder1965Simplex}
Nelder, J.~A., \& Mead, R. 1965, Computer Journal, 7, 308

\bibitem[{{Noguchi} {et~al.}(2002){Noguchi}, {Aoki}, {Kawanomoto}, {Ando},
  {Honda}, {Izumiura}, {Kambe}, {Okita}, {Sadakane}, {Sato}, {Tajitsu},
  {Takada-Hidai}, {Tanaka}, {Watanabe}, \& {Yoshida}}]{2002PASJ...54..855N}
{Noguchi}, K., {Aoki}, W., {Kawanomoto}, S., {et~al.} 2002, \pasj, 54, 855

\bibitem[{{Omiya} {et~al.}(2012){Omiya}, {Han}, {Izumioura}, {Lee}, {Sato},
  {Kim}, {Yoon}, {Kambe}, {Yoshida}, {Masuda}, {Toyota}, {Urakawa}, \&
  {Takada-Hidai}}]{2012PASJ...64...34O}
{Omiya}, M., {Han}, I., {Izumioura}, H., {et~al.} 2012, \pasj, 64, 34

\bibitem[{{Peek} {et~al.}(2009){Peek}, {Johnson}, {Fischer}, {Marcy}, {Henry},
  {Howard}, {Wright}, {Lowe}, {Reffert}, {Schwab}, {Williams}, {Isaacson}, \&
  {Giguere}}]{2009PASP..121..613P}
{Peek}, K.~M.~G., {Johnson}, J.~A., {Fischer}, D.~A., {et~al.} 2009, \pasp,
  121, 613

\bibitem[{{Pepe} {et~al.}(2007){Pepe}, {Correia}, {Mayor}, {Tamuz}, {Couetdic},
  {Benz}, {Bertaux}, {Bouchy}, {Laskar}, {Lovis}, {Naef}, {Queloz}, {Santos},
  {Sivan}, {Sosnowska}, \& {Udry}}]{2007A&A...462..769P}
{Pepe}, F., {Correia}, A.~C.~M., {Mayor}, M., {et~al.} 2007, \aap, 462, 769

\bibitem[{{Perryman} \& {ESA}(1997)}]{1997ESASP1200.....P}
{Perryman}, M.~A.~C., \& {ESA}, eds. 1997, ESA Special Publication, Vol. 1200,
  {The HIPPARCOS and TYCHO catalogues. Astrometric and photometric star
  catalogues derived from the ESA HIPPARCOS Space Astrometry Mission}

\bibitem[{{Pont} {et~al.}(2011){Pont}, {Aigrain}, \&
  {Zucker}}]{2011MNRAS.411.1953P}
{Pont}, F., {Aigrain}, S., \& {Zucker}, S. 2011, \mnras, 411, 1953

\bibitem[{{Press} {et~al.}(1986){Press}, {Flannery}, \&
  {Teukolsky}}]{1986nras.book.....P}
{Press}, W.~H., {Flannery}, B.~P., \& {Teukolsky}, S.~A. 1986, {Numerical
  recipes. The art of scientific computing}

\bibitem[{{Santos} {et~al.}(2003){Santos}, {Israelian}, {Mayor}, {Rebolo}, \&
  {Udry}}]{2003A&A...398..363S}
{Santos}, N.~C., {Israelian}, G., {Mayor}, M., {Rebolo}, R., \& {Udry}, S.
  2003, \aap, 398, 363

\bibitem[{{Santos} {et~al.}(2014){Santos}, {Mortier}, {Faria}, {Dumusque},
  {Adibekyan}, {Delgado-Mena}, {Figueira}, {Benamati}, {Boisse}, {Cunha},
  {Gomes da Silva}, {Lo Curto}, {Lovis}, {Martins}, {Mayor}, {Melo}, {Oshagh},
  {Pepe}, {Queloz}, {Santerne}, {S{\'e}gransan}, {Sozzetti}, {Sousa}, \&
  {Udry}}]{2014A&A...566A..35S}
{Santos}, N.~C., {Mortier}, A., {Faria}, J.~P., {et~al.} 2014, \aap, 566, A35

\bibitem[{{Sato} {et~al.}(2002){Sato}, {Kambe}, {Takeda}, {Izumiura}, \&
  {Ando}}]{2002PASJ...54..873S}
{Sato}, B., {Kambe}, E., {Takeda}, Y., {Izumiura}, H., \& {Ando}, H. 2002,
  \pasj, 54, 873

\bibitem[{{Sato} {et~al.}(2005){Sato}, {Fischer}, {Henry}, {Laughlin},
  {Butler}, {Marcy}, {Vogt}, {Bodenheimer}, {Ida}, {Toyota}, {Wolf}, {Valenti},
  {Boyd}, {Johnson}, {Wright}, {Ammons}, {Robinson}, {Strader}, {McCarthy},
  {Tah}, \& {Minniti}}]{2005ApJ...633..465S}
{Sato}, B., {Fischer}, D.~A., {Henry}, G.~W., {et~al.} 2005, \apj, 633, 465

\bibitem[{{Sato} {et~al.}(2013){Sato}, {Omiya}, {Harakawa}, {Liu}, {Izumiura},
  {Kambe}, {Takeda}, {Yoshida}, {Itoh}, {Ando}, {Kokubo}, \&
  {Ida}}]{2013PASJ...65...85S}
{Sato}, B., {Omiya}, M., {Harakawa}, H., {et~al.} 2013, \pasj, 65, 85

\bibitem[{{Scargle}(1982)}]{1982ApJ...263..835S}
{Scargle}, J.~D. 1982, \apj, 263, 835

\bibitem[{{Tajitsu} {et~al.}(2012){Tajitsu}, {Aoki}, \&
  {Yamamuro}}]{2012PASJ...64...77T}
{Tajitsu}, A., {Aoki}, W., \& {Yamamuro}, T. 2012, \pasj, 64, 77

\bibitem[{{Valenti} \& {Fischer}(2005)}]{2005ApJS..159..141V}
{Valenti}, J.~A., \& {Fischer}, D.~A. 2005, \apjs, 159, 141

\bibitem[{{Valenti} \& {Piskunov}(1996)}]{1996A&AS..118..595V}
{Valenti}, J.~A., \& {Piskunov}, N. 1996, \aaps, 118, 595

\bibitem[{{Vogt} {et~al.}(2005){Vogt}, {Butler}, {Marcy}, {Fischer}, {Henry},
  {Laughlin}, {Wright}, \& {Johnson}}]{2005ApJ...632..638V}
{Vogt}, S.~S., {Butler}, R.~P., {Marcy}, G.~W., {et~al.} 2005, \apj, 632, 638

\bibitem[{{Vogt} {et~al.}(1994){Vogt}, {Allen}, {Bigelow}, {Bresee}, {Brown},
  {Cantrall}, {Conrad}, {Couture}, {Delaney}, {Epps}, {Hilyard}, {Hilyard},
  {Horn}, {Jern}, {Kanto}, {Keane}, {Kibrick}, {Lewis}, {Osborne},
  {Pardeilhan}, {Pfister}, {Ricketts}, {Robinson}, {Stover}, {Tucker}, {Ward},
  \& {Wei}}]{1994SPIE.2198..362V}
{Vogt}, S.~S., {Allen}, S.~L., {Bigelow}, B.~C., {et~al.} 1994, in Society of
  Photo-Optical Instrumentation Engineers (SPIE) Conference Series, Vol. 2198,
  Instrumentation in Astronomy VIII, ed. D.~L. {Crawford} \& E.~R. {Craine},
  362

\bibitem[{{Winn} {et~al.}(2009){Winn}, {Johnson}, {Albrecht}, {Howard},
  {Marcy}, {Crossfield}, \& {Holman}}]{2009ApJ...703L..99W}
{Winn}, J.~N., {Johnson}, J.~A., {Albrecht}, S., {et~al.} 2009, \apjl, 703, L99

\bibitem[{{Wittenmyer} {et~al.}(2014){Wittenmyer}, {Horner}, {Tinney},
  {Butler}, {Jones}, {Tuomi}, {Salter}, {Carter}, {Koch}, {O'Toole}, {Bailey},
  \& {Wright}}]{2014ApJ...783..103W}
{Wittenmyer}, R.~A., {Horner}, J., {Tinney}, C.~G., {et~al.} 2014, \apj, 783,
  103

\bibitem[{{Wright}(2005)}]{2005PASP..117..657W}
{Wright}, J.~T. 2005, \pasp, 117, 657

\bibitem[{{Wright} {et~al.}(2009){Wright}, {Upadhyay}, {Marcy}, {Fischer},
  {Ford}, \& {Johnson}}]{2009ApJ...693.1084W}
{Wright}, J.~T., {Upadhyay}, S., {Marcy}, G.~W., {et~al.} 2009, \apj, 693, 1084

\bibitem[{{Wright} {et~al.}(2011){Wright}, {Fakhouri}, {Marcy}, {Han}, {Feng},
  {Johnson}, {Howard}, {Fischer}, {Valenti}, {Anderson}, \&
  {Piskunov}}]{2011PASP..123..412W}
{Wright}, J.~T., {Fakhouri}, O., {Marcy}, G.~W., {et~al.} 2011, \pasp, 123, 412

\bibitem[{{Yasui} {et~al.}(2010){Yasui}, {Kobayashi}, {Tokunaga}, {Saito}, \&
  {Tokoku}}]{2010ApJ...723L.113Y}
{Yasui}, C., {Kobayashi}, N., {Tokunaga}, A.~T., {Saito}, M., \& {Tokoku}, C.
  2010, \apjl, 723, L113

\bibitem[{{Yi} {et~al.}(2001){Yi}, {Demarque}, {Kim}, {Lee}, {Ree}, {Lejeune},
  \& {Barnes}}]{2001ApJS..136..417Y}
{Yi}, S., {Demarque}, P., {Kim}, Y.-C., {et~al.} 2001, \apjs, 136, 417

\end{thebibliography}

\clearpage

\clearpage

\clearpage

\clearpage

\clearpage

\clearpage

\clearpage

\clearpage

\clearpage

\begin{deluxetable}{lcccc}
\tabletypesize{\normalsize}
\tablecolumns{4}
%\rotate
\tablecaption{Respective Stellar Parameters \label{tbl_str} }
\tablewidth{0pt}
\tablehead{\colhead{Parameter} & \colhead{HD 1605} & \colhead{HD 1666} & \colhead{HD 67087}}
\startdata
$V$ & 7.52 & 8.17 & 8.05\\
$M_V$ & 2.88 & 2.76 & 3.31\\
$B-V$ & 0.96 & 0.53 & 0.53\\
Spectral type & K1 IV & F7 V & F8 V\\
Distance (pc) & 84.6 & 110.6 & 88.8\\
$T_{eff}$ (K) & $4757\pm50$ & $6317\pm44$ & $6330\pm47$\\
$\logg\ (cgi)$ & $3.40\pm0.07$ & $4.06\pm0.06$ & $4.19\pm0.06$\\
$\left[ {\rm Fe/H}\right]$ & $+0.21\pm0.05$ & $+0.37\pm0.04$ & $+0.25\pm0.04$\\
$\vsini\ (\kmps)$ & $0.54\pm0.5$ & $5.6\pm0.5$ & $9.7\pm0.5$\\
$M_*\ (M_\odot )$ & $1.31\pm0.11$ & $1.50\pm0.07$ & $1.36\pm0.04$\\
$R_* (R_\odot )$ & $3.8\pm0.4$ & $1.93\pm0.49$ & $1.55\pm0.32$\\
$L_* (L_\odot )$ & $6.6\pm0.76$ & $5.37\pm1.18$ & $3.47\pm0.75$\\
$B.C.$ & -0.432 & -0.016 & -0.010 \\
age (Gyr) & $4.59\pm1.37$ & $1.76\pm0.20$ & $1.45\pm0.51$
\enddata
\end{deluxetable}

\clearpage
\LongTables
\begin{deluxetable}{rrrc}
\tablecolumns{4}
\tablecaption{RVs for HD 1605. \label{tbl_obs1605}}
\tabletypesize{\normalsize}
\tablewidth{0pt}
\tablehead{
\colhead{Observation Date} & \colhead{RV} & \colhead{Uncertainties} & \colhead{Observatory} \\
\colhead{(JD $- 2450000$)} & \colhead{(${\rm ms^{-1}}$)} & \colhead{(${\rm ms^{-1}}$)} & \colhead{}
}
\startdata
3242.026396 & 55.28 & 3.85 & Subaru \\ 
3360.824924 & 74.05 & 3.47 & Subaru \\ 
3361.734252 & 68.41 & 3.59 & Subaru \\ 
3401.755798 & 60.49 & 3.69 & Subaru \\ 
3693.825842 & 6.87 & 3.37 & Subaru \\ 
3726.799867 & 8.19 & 3.60 & Subaru \\ 
3727.684276 & 5.18 & 3.59 & Subaru \\ 
3727.742699 & 4.19 & 3.36 & Subaru \\ 
3727.823795 & -5.14 & 3.56 & Subaru \\ 
3746.791000 & 17.27 & 0.99 & Keck \\ 
3747.788000 & 2.04 & 1.09 & Keck \\ 
3748.716000 & 12.47 & 1.02 & Keck \\ 
3749.719000 & 2.58 & 1.02 & Keck \\ 
3750.706000 & 4.03 & 1.04 & Keck \\ 
3775.734000 & 17.89 & 1.03 & Keck \\ 
3776.712000 & 6.55 & 1.06 & Keck \\ 
4427.814000 & -48.43 & 1.07 & Keck \\ 
4428.839000 & -50.16 & 1.09 & Keck \\ 
4635.073000 & -28.65 & 1.01 & Keck \\ 
4638.113000 & -33.57 & 0.99 & Keck \\ 
4641.098000 & -33.98 & 1.10 & Keck \\ 
4743.946902 & -40.84 & 3.06 & Subaru \\ 
4786.941025 & -51.52 & 3.48 & Subaru \\ 
5044.113000 & 12.41 & 1.06 & Keck \\ 
5056.151530 & 32.23 & 4.63 & OAO/S \\ 
5082.147000 & 19.70 & 1.01 & Keck \\ 
5099.086084 & 32.55 & 3.62 & OAO/S \\ 
5130.997159 & 46.44 & 4.90 & OAO/S \\ 
5159.174942 & 40.19 & 4.61 & OAO/S \\ 
5171.854000 & 36.05 & 1.07 & Keck \\ 
5184.968137 & 44.59 & 3.66 & OAO/S \\ 
5186.975068 & 37.95 & 4.70 & OAO/S \\ 
5187.791000 & 40.23 & 1.07 & Keck \\ 
5234.905796 & 24.64 & 15.17 & OAO/S \\ 
5379.092000 & 3.62 & 1.09 & Keck \\ 
5398.230411 & 15.43 & 4.54 & OAO/S \\ 
5401.227137 & 14.15 & 7.79 & OAO/S \\ 
5402.233880 & 7.37 & 4.29 & OAO/S \\ 
5404.234587 & 7.69 & 4.67 & OAO/S \\ 
5409.228767 & 15.62 & 3.72 & OAO/S \\ 
5410.204903 & 14.99 & 4.46 & OAO/S \\ 
5428.071809 & 17.37 & 3.22 & OAO/F \\ 
5429.125070 & 17.21 & 3.22 & OAO/F \\ 
5430.151899 & 20.53 & 3.14 & OAO/F \\ 
5431.302805 & 13.18 & 3.05 & OAO/F \\ 
5432.148032 & 17.45 & 3.20 & OAO/F \\ 
5433.162621 & 15.67 & 3.53 & OAO/F \\ 
5434.218731 & 13.32 & 3.45 & OAO/F \\ 
5435.173909 & 13.84 & 3.06 & OAO/F \\ 
5441.181714 & 3.31 & 3.11 & OAO/S \\ 
5445.266316 & 4.96 & 4.13 & OAO/S \\ 
5468.117129 & 0.50 & 4.45 & OAO/S \\ 
5474.112849 & 3.61 & 3.55 & OAO/S \\ 
5476.138541 & 1.69 & 3.52 & OAO/S \\ 
5480.204613 & -2.43 & 4.52 & OAO/S \\ 
5502.094646 & 5.50 & 4.11 & OAO/S \\ 
5506.089555 & 1.85 & 3.29 & OAO/S \\ 
5525.033422 & 3.43 & 3.85 & OAO/S \\ 
5545.010623 & 7.29 & 5.98 & OAO/S \\ 
5546.990257 & 18.82 & 5.16 & OAO/S \\ 
5570.972884 & 27.02 & 4.78 & OAO/F \\ 
5596.902826 & 25.46 & 2.98 & OAO/F \\ 
5618.910983 & 25.10 & 3.27 & OAO/F \\ 
5696.303752 & 47.87 & 4.20 & OAO/F \\ 
5697.295470 & 39.86 & 6.77 & OAO/F \\ 
5700.291167 & 25.83 & 3.38 & OAO/F \\ 
5701.292900 & 30.94 & 3.53 & OAO/F \\ 
5757.160078 & 30.11 & 3.85 & OAO/S \\ 
5758.277612 & 11.75 & 4.92 & OAO/S \\ 
5759.162776 & 22.78 & 4.76 & OAO/S \\ 
5772.170736 & 29.60 & 7.17 & OAO/F \\ 
5772.212740 & 19.76 & 2.98 & OAO/F \\ 
5781.135733 & 5.52 & 3.63 & Subaru \\ 
5812.030734 & 8.19 & 2.79 & OAO/F \\ 
5813.038857 & 8.60 & 2.64 & OAO/F \\ 
5814.041536 & 11.69 & 2.97 & OAO/F \\ 
5817.074067 & 9.19 & 2.99 & OAO/F \\ 
5842.081940 & -4.83 & 2.94 & OAO/F \\ 
5843.009333 & -2.09 & 3.07 & OAO/F \\ 
5910.995010 & -32.82 & 3.77 & OAO/F \\ 
5931.934054 & -35.43 & 2.65 & OAO/F \\ 
6132.215925 & -50.30 & 5.02 & OAO/F \\ 
6135.173177 & -40.70 & 3.73 & OAO/F \\ 
6138.167379 & -43.62 & 4.32 & OAO/F \\ 
6148.110402 & -40.38 & 3.13 & Subaru/IS\#1 \\ 
6149.102357 & -48.86 & 4.70 & Subaru/IS\#1 \\ 
6163.122505 & -38.39 & 5.08 & OAO/F \\ 
6240.975244 & -41.49 & 3.91 & OAO/F \\ 
6279.939767 & -42.60 & 3.49 & OAO/F \\ 
6514.129844 & -98.55 & 3.97 & OAO/F \\ 
6517.149862 & -96.22 & 3.50 & OAO/F \\ 
6523.128885 & -100.38 & 3.62 & OAO/F \\ 
\enddata
\tablecomments{OAO/S and OAO/F represent data observed from OAO 188 cm telescope with HIDES-Slit mode and HIDES-Fiber feed mode, respectively.
The RV data from Subaru/HDS using IS\#1 are presented as Subaru/IS\#1.
All RV values are subtracted by appropriate $\gamma$ value listed in Table \ref{tbl_prm}.
Subsequent RV tables follow this manner.}
\end{deluxetable}

\clearpage
\LongTables
\begin{deluxetable}{rrrc}
\tablecolumns{4}
\tablecaption{RVs for HD 1666. \label{tbl_obs1666} }
\tabletypesize{\normalsize}
\tablewidth{0pt}
\tablehead{
\colhead{Observation Date} & \colhead{RV} & \colhead{Uncertainties} & \colhead{Observatory} \\
\colhead{(JD $- 2450000$)} & \colhead{(${\rm ms^{-1}}$)} & \colhead{(${\rm ms^{-1}}$)} & \colhead{}
}
\startdata
3360.701854 & 34.89 & 6.71 & Subaru \\ 
3361.693030 & 11.01 & 7.51 & Subaru \\ 
3362.692462 & 26.71 & 6.80 & Subaru \\ 
3401.709260 & -80.49 & 9.41 & Subaru \\ 
4078.817044 & 281.88 & 6.86 & Subaru \\ 
4080.798905 & 257.23 & 7.26 & Subaru \\ 
4081.784028 & 237.31 & 5.83 & Subaru \\ 
4743.990920 & -80.56 & 8.37 & Subaru \\ 
5099.134235 & -139.93 & 11.31 & OAO/S \\ 
5131.047738 & -125.16 & 16.27 & OAO/S \\ 
5184.894363 & 159.10 & 12.13 & OAO/S \\ 
5186.903685 & 168.04 & 11.78 & OAO/S \\ 
5396.121716 & -191.64 & 1.72 & Keck \\ 
5397.063226 & -201.03 & 1.64 & Keck \\ 
5400.127388 & -170.83 & 1.76 & Keck \\ 
5401.051892 & -150.05 & 2.00 & Keck \\ 
5401.274669 & -126.25 & 30.18 & OAO/S \\ 
5402.283521 & -141.41 & 14.87 & OAO/S \\ 
5403.104754 & -108.01 & 1.58 & Keck \\ 
5404.282251 & 52.41 & 20.70 & OAO/S \\ 
5405.092260 & -19.88 & 1.73 & Keck \\ 
5406.088778 & -53.29 & 1.67 & Keck \\ 
5407.107137 & -68.75 & 1.62 & Keck \\ 
5410.287100 & -11.85 & 21.87 & OAO/S \\ 
5411.112017 & -8.15 & 1.56 & Keck \\ 
5412.006057 & -20.98 & 1.62 & Keck \\ 
5413.071693 & 19.40 & 1.58 & Keck \\ 
5414.079847 & 91.11 & 1.64 & Keck \\ 
5426.081973 & 295.80 & 1.95 & Keck \\ 
5428.102646 & 286.88 & 1.85 & Keck \\ 
5429.100236 & 272.53 & 1.81 & Keck \\ 
5429.220400 & 242.83 & 14.74 & OAO/F \\ 
5430.199245 & 219.08 & 14.58 & OAO/F \\ 
5431.210050 & 211.42 & 12.46 & OAO/F \\ 
5432.241269 & 177.79 & 14.56 & OAO/F \\ 
5433.243331 & 129.94 & 13.56 & OAO/F \\ 
5433.994463 & 224.60 & 1.82 & Keck \\ 
5437.132373 & 241.94 & 1.83 & Keck \\ 
5439.129924 & 226.52 & 1.75 & Keck \\ 
5440.044256 & 191.75 & 1.83 & Keck \\ 
5442.191536 & 203.51 & 12.40 & OAO/S \\ 
5444.192803 & 198.81 & 12.47 & OAO/S \\ 
5445.188846 & 187.01 & 11.86 & OAO/S \\ 
5455.950750 & 124.52 & 1.96 & Keck \\ 
5465.039075 & 148.60 & 1.84 & Keck \\ 
5471.108357 & 98.93 & 11.72 & OAO/S \\ 
5475.112576 & 133.07 & 14.14 & OAO/S \\ 
5476.113014 & 131.94 & 13.30 & OAO/S \\ 
5480.084970 & 56.51 & 17.92 & OAO/S \\ 
5503.986138 & 52.05 & 11.18 & OAO/S \\ 
5525.008388 & -1.55 & 10.30 & OAO/S \\ 
5548.949071 & 2.94 & 11.34 & OAO/S \\ 
5556.925781 & -52.35 & 15.54 & OAO/S \\ 
5570.892782 & -30.05 & 15.01 & OAO/F \\ 
5758.302148 & 139.44 & 12.77 & OAO/S \\ 
5759.268807 & 75.19 & 15.06 & OAO/S \\ 
5780.140654 & 31.75 & 7.99 & Subaru \\ 
5812.211520 & -25.19 & 9.26 & OAO/F \\ 
5813.142636 & -15.14 & 12.14 & OAO/F \\ 
5814.175882 & 18.86 & 12.09 & OAO/F \\ 
5816.172657 & -9.86 & 11.06 & OAO/F \\ 
5817.137730 & -56.90 & 9.44 & OAO/F \\ 
5842.058927 & -49.84 & 13.19 & OAO/F \\ 
5843.132379 & -70.46 & 10.49 & OAO/F \\ 
5845.097980 & -52.67 & 10.81 & OAO/F \\ 
5845.119053 & -48.87 & 9.46 & OAO/F \\ 
5846.094836 & -31.20 & 10.73 & OAO/F \\ 
5846.116016 & -16.82 & 11.20 & OAO/F \\ 
5901.014322 & -59.08 & 15.15 & OAO/F \\ 
5901.963414 & -65.42 & 11.41 & OAO/F \\ 
5903.960129 & -170.84 & 11.45 & OAO/F \\ 
5904.963532 & -157.23 & 9.10 & OAO/F \\ 
5905.999945 & -153.12 & 10.62 & OAO/F \\ 
5907.960761 & -125.94 & 8.91 & OAO/F \\ 
5909.960182 & -125.27 & 11.24 & OAO/F \\ 
5910.968214 & -99.10 & 17.90 & OAO/F \\ 
5911.934768 & -76.82 & 10.49 & OAO/F \\ 
5912.892384 & -85.77 & 7.29 & OAO/F \\ 
5912.913863 & -91.26 & 8.30 & OAO/F \\ 
5913.903891 & -120.54 & 21.31 & OAO/F \\ 
5914.901327 & -168.36 & 7.27 & OAO/F \\ 
5914.921310 & -170.06 & 13.28 & OAO/F \\ 
5930.879865 & -159.39 & 11.11 & OAO/F \\ 
5931.877692 & -182.01 & 8.77 & OAO/F \\ 
5932.889575 & -134.12 & 8.41 & OAO/F \\ 
5933.885456 & -99.03 & 8.10 & OAO/F \\ 
5934.878720 & -102.03 & 8.05 & OAO/F \\ 
5935.888159 & -125.08 & 12.19 & OAO/F \\ 
6134.296666 & -117.54 & 13.65 & OAO/F \\ 
6136.265378 & -90.49 & 14.10 & OAO/F \\ 
6136.286852 & -88.34 & 12.79 & OAO/F \\ 
6148.092136 & -78.26 & 7.32 & Subaru/IS\#1 \\ 
6149.066180 & -102.46 & 6.81 & Subaru/IS\#1 \\ 
6164.265986 & -101.72 & 10.27 & OAO/F \\ 
6165.246902 & -69.81 & 10.73 & OAO/F \\ 
6241.051622 & 314.86 & 15.36 & OAO/F \\ 
6274.952788 & 113.00 & 7.67 & OAO/F \\ 
6281.899790 & 95.35 & 8.57 & OAO/F \\ 
6525.307091 & 157.21 & 10.77 & OAO/F \\ 
\enddata
\end{deluxetable}

\clearpage

\begin{deluxetable}{rrrc}
\tablecolumns{4}
\tablecaption{RVs for HD 67087.}
\tabletypesize{\normalsize}
\tablewidth{0pt}
\tablehead{
\colhead{Observation Date} & \colhead{RV} & \colhead{Uncertainties} & \colhead{Observatory} \\
\colhead{(JD $- 2450000$)} & \colhead{(${\rm ms^{-1}}$)} & \colhead{(${\rm ms^{-1}}$)} & \colhead{}
}
\startdata
3361.119675 & 85.64 & 11.26 & Subaru \\ 
4079.034701 & 70.38 & 11.03 & Subaru \\ 
4080.041048 & 61.08 & 8.87 & Subaru \\ 
4081.107316 & 63.50 & 10.11 & Subaru \\ 
4523.858969 & -27.12 & 10.13 & Subaru \\ 
4587.797606 & -81.32 & 9.86 & Subaru \\ 
4744.134303 & -31.62 & 10.15 & Subaru \\ 
4787.077253 & 13.78 & 9.21 & Subaru \\ 
5566.282107 & 46.39 & 13.39 & OAO/F \\ 
5591.177854 & 26.16 & 16.90 & OAO/F \\ 
5592.185199 & 43.64 & 13.45 & OAO/F \\ 
5594.176621 & 46.84 & 17.35 & OAO/F \\ 
5595.178250 & 37.70 & 13.91 & OAO/F \\ 
5596.178941 & 17.20 & 11.44 & OAO/F \\ 
5597.200501 & 42.93 & 13.36 & OAO/F \\ 
5619.140168 & -7.08 & 15.94 & OAO/F \\ 
5649.114480 & -15.63 & 12.04 & OAO/F \\ 
5651.125157 & -2.28 & 13.26 & OAO/F \\ 
5655.035694 & -21.83 & 18.42 & OAO/F \\ 
5667.974503 & -53.30 & 21.34 & OAO/F \\ 
5669.008561 & -33.89 & 14.46 & OAO/F \\ 
5671.004736 & -17.33 & 18.93 & OAO/F \\ 
5672.014501 & -26.10 & 12.13 & OAO/F \\ 
5695.003326 & -31.91 & 14.75 & OAO/F \\ 
5695.960308 & -56.16 & 13.23 & OAO/F \\ 
5699.969525 & -39.14 & 14.30 & OAO/F \\ 
5700.984984 & -48.07 & 16.30 & OAO/F \\ 
5815.320048 & 71.48 & 17.03 & OAO/F \\ 
5816.321979 & 84.62 & 17.37 & OAO/F \\ 
5842.301613 & 80.10 & 11.41 & OAO/F \\ 
5843.336659 & 73.52 & 12.64 & OAO/F \\ 
5845.304024 & 93.58 & 15.53 & OAO/F \\ 
5846.320505 & 94.84 & 12.42 & OAO/F \\ 
5854.102996 & 92.03 & 8.96 & Subaru/IS\#1 \\ 
5904.270617 & 67.56 & 14.79 & OAO/F \\ 
5911.129521 & 78.21 & 14.21 & OAO/F \\ 
5932.040732 & 30.38 & 13.50 & OAO/F \\ 
5965.031544 & 5.15 & 8.41 & Subaru \\ 
5966.079041 & -22.73 & 17.53 & OAO/F \\ 
5967.932216 & 0.41 & 9.97 & Subaru \\ 
5968.821621 & -7.63 & 9.58 & Subaru \\ 
6014.031330 & -44.49 & 11.49 & OAO/F \\ 
6015.034135 & -51.38 & 12.28 & OAO/F \\ 
6040.960413 & -77.30 & 12.72 & OAO/F \\ 
6057.947730 & -86.41 & 16.10 & OAO/F \\ 
6241.358500 & 41.08 & 15.18 & OAO/F \\ 
6275.300959 & 29.48 & 12.55 & OAO/F \\ 
6554.286423 & 61.06 & 13.23 & OAO/F \\ 
6758.939170 & -105.20 & 11.07 & OAO/F \\ 
6780.966475 & -85.21 & 11.18 & OAO/F \\ 
6784.989355 & -115.59 & 13.17 & OAO/F \\ 
\enddata
\label{tbl_obs67087}
\end{deluxetable}

\clearpage

\begin{deluxetable}{lccccc}
\tabletypesize{\small}
\tablecolumns{6}
%\rotate
\tablecaption{Orbital Solutions for the newly detected planetary systems \label{tbl_prm}}
\tablewidth{0pt}
\tablehead{\colhead{Parameter} & \colhead{HD 1605 b} & \colhead{HD 1605 c} & \colhead{HD 1666 b} & \colhead{HD 67087 b} & \colhead{HD 67087 c}}
\startdata
$P$(d) & 
 $577.9^{+5.6}_{-4.9}$ & $2111^{+42}_{-32}$ & 
 $270.0^{+0.8}_{-0.9}$ & 
 $352.2^{+1.7}_{-1.6}$ & $2374^{+193}_{-156}$\\
$T_c$(JD$-2450000$) & 
 $3443.3^{+58.9}_{-67.3}$ & $4758.3^{+105.1}_{-110.0}$ & 
 $3526.3^{+7.3}_{-5.3}$ & 
 $154.8^{+26.3}_{-29.1}$ & $322.5^{+370.6}_{-407.1}$\\
$e$ & 
 $0.078^{+0.082}_{-0.035}$ & $0.098^{+0.032}_{-0.022}$ & 
 $0.63^{+0.03}_{-0.02}$ & 
 $0.17^{+0.07}_{-0.07}$ & $0.76^{+0.17}_{-0.24}$\\
$\omega$(rad) & 
 $0.45^{+0.65}_{-0.75}$ & $4.20^{+0.29}_{-0.30}$ & 
 $5.13^{+0.09}_{-0.01}$ & 
 $4.97^{+0.36}_{-0.29}$ & $4.46^{+0.48}_{-0.43}$\\
$K_1$($\mps$) & 
 $19.8^{+1.2}_{-0.8}$ & $46.5^{+1.5}_{-1.2}$ & 
 $199.4^{+8.8}_{-5.5}$ & 
 $73.6^{+4.4}_{-3.9}$ & $93.3^{+151.4}_{-40.8}$
 \\
$a_1\sin i$($10^{-3}$AU) & 
 $1.05^{+0.06}_{-0.05}$ & $8.98^{+0.35}_{-0.26}$ & 
 $3.83^{+0.13}_{-0.08}$ & 
 $2.35^{+0.18}_{-0.16}$ & $13.4^{+28.5}_{-10.8}$\\
$a_{\rm p}$(AU) & 
 $1.48^{+0.02}_{-0.01}$ & $3.52^{+0.05}_{-0.05}$ & 
 $0.94^{+0.01}_{-0.02}$ & 
 $1.08^{+0.04}_{-0.04}$ & $3.86^{+0.43}_{-0.37}$\\
$\msini$($\mjup$) & 
 $0.96^{+0.06}_{-0.04}$ & $3.48^{+0.13}_{-0.11}$ & 
 $6.43^{+0.31}_{-0.22}$ & 
 $3.06^{+0.22}_{-0.20}$ & $4.85^{+10.0}_{-3.61}$
 \\
Trend($\mps{\rm yr^{-1}}$) & 
 \multicolumn{2}{c}{$-4.92^{+0.17}_{-0.22}$} & 
 \nodata & 
 \multicolumn{2}{c}{\nodata}\\
$N_{\rm obs}$ & 
 \multicolumn{2}{c}{$92$} & 
 $99$ & 
 \multicolumn{2}{c}{$51$}\\
$\sigma_{\rm jitt}$ $(\mps)$ & 
 \multicolumn{2}{c}{6.0} & 
 34.6 & 
 \multicolumn{2}{c}{4.0}\\
rms($\mps$) & 
 \multicolumn{2}{c}{$6.4$} & 
 $35.6$ & 
 \multicolumn{2}{c}{$11.8$}\\
$\sqrt{\chi_\nu^2}$ & 
 \multicolumn{2}{c}{$0.97$} & 
 1.00 & 
 \multicolumn{2}{c}{0.90\tablenotemark{a}}\\
$\gamma_\mathrm{HDS \& HIDES}$ $(\mps)$ & 
 \multicolumn{2}{c}{$6.7$} & 
 $-7.4$ &
 \multicolumn{2}{c}{$17.0$}\\
$\gamma_\mathrm{HIRES}$ $(\mps)$ & 
 \multicolumn{2}{c}{$0.1$} & 
 $54.3$ &
 \multicolumn{2}{c}{\nodata}
\enddata
\tablenotetext{a}{The best-fit parameters from the MCMC-fit is applied.}
\end{deluxetable}

\clearpage

\end{document}